\documentclass[final,1p,times,number]{elsarticle}

\usepackage{hyperref}
\usepackage{tikz}
\usetikzlibrary{arrows.meta,calc,decorations.markings,math,arrows.meta,shapes.arrows, fadings}
\usepackage{physics}
\usepackage{tikz-feynman, amsmath}
\graphicspath{{img/}}
\usepackage{comment}
\usepackage{float}
\usepackage{units}
\usepackage{amssymb}
\usepackage{xcolor}

\usepackage{subcaption}
\usepackage{caption}

\definecolor{darkblue}{rgb}{0.1,0.2,0.6}

\newcommand{\be}{ \begin{equation} }
\newcommand{\ee}{\end{equation}}
\newcommand{\bea}{ \begin{eqnarray} }
\newcommand{\eea}{\end{eqnarray}}

\journal{Annals of Physics}

\begin{document}

\begin{frontmatter}

\title{Dynamics of Many-Body Delocalization in the Time-dependent Hartree-Fock Approximation} 


\author{Paul P\"opperl\footnote{paul.poepperl@kit.edu}, Elmer V.~H.~Doggen, Jonas F.~Karcher}
\address{{Institute for Quantum Materials and Technologies, Karlsruhe Institute of Technology, 76021 Karlsruhe, Germany}}
\address{{Institut f\"ur Theorie der Kondensierten Materie, Karlsruhe Institute of Technology, 76128 Karlsruhe, Germany}}
\author{\mbox{Alexander~D.~Mirlin}}
\address{{Institute for Quantum Materials and Technologies, Karlsruhe Institute of Technology, 76021 Karlsruhe, Germany}}
\address{{Institut f\"ur Theorie der Kondensierten Materie, Karlsruhe Institute of Technology, 76128 Karlsruhe, Germany}}
\address{L.~D. Landau Institute for Theoretical Physics RAS, 119334 Moscow, Russia}
\address{Petersburg Nuclear Physics Institute, 188300 St.~Petersburg, Russia}
\author{Konstantin S.~Tikhonov}
\address{Skolkovo Institute of Science and Technology, Moscow, 121205, Russia}
\address{Condensed-matter Physics Laboratory, National Research University Higher School of Economics, 101000 Moscow, Russia}

\begin{abstract}

We explore dynamics of disordered and quasi-periodic interacting lattice models using a self-consistent time-dependent Hartree-Fock (TDHF) approximation, accessing both large systems (up to $L = 400$ sites) and very long times (up to $t = 10^5$). We find that, in the $t \to \infty$ limit, the many-body localization (MBL) is always destroyed within the TDHF approximation. At the same time, this approximation provides important information on the long-time character of dynamics in the ergodic side of the MBL transition. Specifically, for one-dimensional (1D) disordered chains, we find slow power-law transport up to the longest times, supporting the rare-region (Griffiths) picture.  The information on this subdiffusive dynamics is obtained by the analysis of three different observables--- temporal decay $\sim t^{-\beta}$ of real-space and energy-space imbalances as well as domain wall melting---which all yield consistent results. For two-dimensional (2D) systems, the decay is faster than a power law, in consistency with theoretical predictions that $\beta$ grows as $\log t$ for the decay governed by rare regions. At longest times and moderately strong disorder, $\beta$ approaches the limiting value $\beta=1$ corresponding to 2D diffusion. In quasi-periodic (Aubry-Andr\'e) 1D systems, where rare regions are absent, we find considerably faster decay that reaches the ballistic value $\beta=1$, which provides further support to the Griffiths picture of the slow transport in random systems.

\end{abstract}


\end{frontmatter}

\section{Introduction}

Many-body localization (MBL) concerns the interplay of disorder and interactions in highly excited states of a many-body system. It is understood that a disordered system that would be Anderson localized in the absence of interaction \cite{paper:absence_of_diffusion} remains localized upon the introduction of interactions in a certain range of parameters \cite{ paper:mbl_gmp, paper:mbl_baa,Altman2015a, Nandkishore2015a, Abanin2017a, Alet2018a,abanin2019colloquium}. The transition between the ergodic and MBL phases is called the MBL transition. 
For one-dimensional (1D) systems with short-range interactions, the MBL transition is believed to exist in the thermodynamic limit at fixed (system-size-independent) values of disorder and interaction. In other situations (including higher-dimensional systems and models with long-range interactions), the MBL transition still exists but in a more sophisticated sense, requiring appropriate scaling of the disorder with the system size. 

The physics of the MBL transition and of both phases surrounding it is of great interest. Remarkably, not only the MBL phase is highly unusual but also the delocalized phase exhibits distinct and unconventional properties. Specifically, it has been found numerically that the ergodic side of the MBL transition features slow, subdiffusive transport in a broad range of parameters \cite{paper:mbl_powerlaws,gopalakrishnan2020dynamics}. The nature of this slow dynamics is still under debate. In particular, it was proposed that the slow transport is due to rare regions of anomalously high disorder leading to Griffiths-type physics. 

The problem that one deals with in this context---a many-body problem involving both disorder and interactions---turns out to be notoriously difficult, both for analytical studies and numerical simulations. While a number of analytical approaches to the MBL problem have been developed, all of them either involve approximations or assumptions that are difficult to control, or treat some simplified toy models. In this situation, numerical simulations are of particular importance. However, the computational treatment of the MBL problem is also an extremely challenging problem.
An exact numerical study of even the simplest quantum many-body systems---1D lattice systems with a single, binary degree of freedom per lattice site---requires a computational complexity that scales as $2^N$, where $N$ is the number of sites in the lattice. While clever numerical techniques have been employed to push this as far as possible \cite{paper:mbl_ed}, this exponential scaling of exact diagonalization (ED) methods is unavoidable, and systems beyond $N \approx 24$ are inaccessible by such methods using reasonable computational resources. While for simpler physical phenomena systems of this size might be already reasonably sufficient, they are much too small for understanding the large-$N$ physics of high-energy many-body states of strongly disordered interacting systems. Thus, alternative computational approaches that give access to long-time dynamics in much larger systems are of vital importance.

 A powerful framework for simulating the dynamics of large quantum many-body systems involves methods based on matrix product states (MPS) and related approaches, such as the time-dependent density matrix renormalization group (t-DMRG) \cite{Daley2004a, White2004a}, time-evolving block decimation (TEBD) \cite{Vidal2003a, Vidal2004a} and the time-dependent variational principle (TDVP) \cite{paper:TDVP_1, paper:TDVP_2} (for a recent review of MPS-based methods for dynamics, see Ref.~\cite{Paeckel2019a}). These methods have been applied to investigations of the MBL problem in several works, including Refs.~\cite{Bardarson2012a, Lim2016a, Prelovsek2016a, Sierant2017a, Sierant2018a, Zakrzewski2018a, Kloss2018a, paper:mbl_elmer, Doggen2019a, paper:elmer_2d, Chanda2020a}, see also the review \cite{mps_mbl_review} and references therein. In particular, the TDVP approach has allowed  to explore the drift of the MBL transition point $W_c(L)$ with the length $L$ for 1D chains \cite{paper:mbl_elmer}. It was found that, while this drift is rather substantial between $L \approx 20$ (available for ED) and $L=50$, it saturates at $L \approx 50$ --- 100, supporting the existence of the MBL transition at finite $W_c$ in the limit $L \to \infty$ in 1D. Further, the TDVP approach has 
demonstrated a qualitative difference between 1D and 2D problems in this respect: in the 2D case, the results support an increase of  $W_c$ with the system size without bounds, as expected from the avalanche theory \cite{paper:elmer_2d}. At the same time, the TDVP approach has its limitations. Specifically, for the large systems, it is limited by relatively modest times (typically hundreds of inverse hopping matrix elements) in the vicinity of the MBL transition. For weaker disorder, the limitation is even stronger in view of a fast growth of the entanglement. While these times are of the same order as those probed in many experiments, there are important questions with respect to the behavior at considerably longer times. As an example, Ref.~\cite{Doggen2019a} has provided some evidence of the difference in the long-time dynamics of random and quasi-periodic systems---which is of fundamental importance for understanding the role of rare events. However, it was clear that data for substantially longer times $t$ are needed to verify the significance of the observed trend. 

It is therefore important to explore further computational approaches  to the physics around the MBL transition that may help to access simultaneously large systems (with $N \approx 100$ sites and larger) and long times, $ t > 10^3$.  Such approaches necessarily involve additional approximations, and one has to investigate whether they provide at least qualitative (or, perhaps, semi-quantitative) understanding of the relevant physics.  Various approaches to quantum dynamics in this type of systems have been suggested in recent years \cite{paper:scb,paper:cluster_truncated_wigner,DeTomasi2019a,Sajna_2020,memories_of_initial_states}.
In the present paper, we focus on the time-dependent Hartree-Fock approximation (TDHF), which has been put forward as a potentially fruitful method for describing MBL systems 
in the recent paper \cite{paper:hf_original}. The TDHF is exact in the non-interacting limit and treats the interaction self-consistently.

The goal of this work is to explore systematically the dynamics in MBL systems within the TDHF approach. We first compare the approach to state-of-the-art ``exact'' methods such as ED and the TDVP (for system sizes and times accessible to these methods). While we do observe clear deviations, we see that the TDHF approach does capture the key ingredient of the problem---which is in the focus of our work---the slow dynamics in the ergodic phase. We thus proceed and perform TDHF numerical simulations up to very long times, $t \sim 10^5$, with system sizes up to 400 sites. 
To study the slow, subdiffusive dynamics of many-body delocalization at long times, we use three different observables: (i) temporal decay of real-space imbalance, (ii) decay of energy-space imbalance, and (iii) melting of a domain wall. Importantly, all these methods yield consistent results for the flowing (time-dependent) power-law exponent $\beta$ that we use to characterize the numerical data.  

One of the central questions that we address is a comparison between the dynamics for random and quasi-periodic 1D systems. We find qualitatively different behavior of $\beta(t)$ in these two cases: while $\beta(t)$ remains relatively small and experiences long-time saturation in random systems, it crosses over to the ballistic value $\beta=1$ in quasi-periodic systems. This provides clear support to the Griffiths (rare-region) mechanism of slow dynamics in random systems; the mechanism that is not operative in quasi-periodic systems \cite{Gopalakrishnan2016a,paper:rare_regions_review}. 

To elucidate the role of spatial dimensionality for MBL, we consider  also two-dimensional (2D) systems. Our results show that the exponent $\beta$ in this case does not saturate at a subdiffusive value but rather grows as $\beta(t) \sim \log t$, again in consistency with expectations based on the Griffiths mechanism of slow transport.  

Our results thus show that the TDHF approach characterizes remarkably well the ergodic side of the MBL transition. A natural and important question to ask is whether the MBL phase (and the MBL transition) are also captured by this approximation. This question is also addressed in the present work. We find (at variance with a proposal in Ref.~\cite{paper:hf_original}) that the MBL phase is always destroyed, in the $t \rightarrow \infty$ limit within the TDHF approximation. The corresponding numerical results are supported by analytical arguments. 

While this manuscript was in preparation, we learned of a related work by Nandy \emph{et al.}~\cite{nandy2020dephasing}, where the TDHF approximation is used, in combination with numerically exact solution, for the analysis of MBL in relatively small 1D systems. Our findings about the absence of a true MBL transition in the TDHF approximation in 1D is consistent with the results of Ref.~\cite{nandy2020dephasing}.

\section{Model and Method}
\label{sec:methods}

\subsection{Spinless Fermi-Hubbard model with on-site potential}
\label{sec:model}

We consider a lattice model of interacting spinless fermions described by the Hamiltonian
\begin{align}
    H_\mathrm{FH} &= \sum_{i=1}^N h_i n_i - J\sum_{i, j=1}^N \delta_{\langle i, j \rangle} c_i^\dagger c_j + U \sum_{i, j=1}^N \delta_{\langle i, j \rangle} n_i n_j \,,\label{eq:h_fermihubbard}\\
    n_i &= c_i^\dagger c_i \,,
\end{align}
where the \(c^\dagger_i\) and \(c_i\) are fermionic creation and annihilation operators in site space and \(N\) is the number of sites. Further, \(\delta_{\langle i, j\rangle}\) is equal to unity if the sites \(i, j\) are nearest neighbors on the considered lattice and zero otherwise. We consider two different types of on-site fields \(h_i\), \(i \in [1, N]\). The first one is \textbf{random disorder}: \(h_i\) are taken as random energies uniformly distributed in the interval \([-W, W]\), yielding an interacting Anderson model. The second case is a \textbf{quasi-periodic} field, leading to an interacting Aubry-Andr\'{e} model:
\begin{align}
    h_i &= \frac{W}{2} \cos(2 \pi \Phi i + \phi_0) \,, \\
    \phi_0 &\in [0, 2 \pi). \label{phi-0}
\end{align}
Here \(\Phi\) is an irrational number rendering the period of the potential incommensurate with the lattice; in this paper we choose \(\Phi=(\sqrt{5} - 1)/2\). Further, \(\phi_0 \in [0,2\pi) \) is a constant phase shift that is taken as a random number over which the averaging is performed. In one dimension, the model \eqref{eq:h_fermihubbard} maps by the Jordan-Wigner transformation to a spin-$1/2$ Hamiltonian. 
We will use the term ``disorder strength'' also for the strength $W$ of the quasi-periodic potential as defined in Eq.~\eqref{phi-0}.
%
%
%

\subsection{Time-dependent Hartree-Fock approximation}

We employ the time-dependent Hartree-Fock (TDHF) approximation to obtain an equation of motion for the lesser Green's function
\begin{align}
    G^<_{i, j}(t, t^\prime) = \mathrm{i} \Tr\Big[ \rho_0 c_j^\dagger(t^\prime) c_i(t) \Big] \,,
\end{align}
where \(\rho_0\) is the density operator corresponding to the initial state and the time-dependent operators are in the Heisenberg picture. Given a Hamiltonian of the form
\begin{align}
    H &= H_0 + \sum_{i, j=1}^N V_{i, j} n_i n_j \,, \\
    H_0 & = \sum_{i=1}^N \varepsilon_i n_i + \sum_{i, j=1}^N J_{i, j} c^\dagger_i c_j \,,
\end{align}
the TDHF equation of motion for \(G_{i, j}^<(t, t^\prime)\) reads
\be
    \mathrm{i} \partial_t \hat{G}^<(t, t^\prime) = \left[ \hat{H}_0 - \hat{\Sigma}^\mathrm{HF}(t) \right] * \hat{G}^<(t, t^\prime) \,,
     \label{eq:hf_eom}
 \ee
    where  \(\Sigma_{i, j}^\mathrm{HF}\) is the Hartree-Fock self-energy,
 \be
    \Sigma_{i, j}^\mathrm{HF}(t) = - \mathrm{i} \delta_{i, j} \sum_k V_{i, k} G_{k, k}^< (t, t) + \mathrm{i} V_{i, j} G_{i, j}^<(t, t) \,.
\ee
 In Eq.~\eqref{eq:hf_eom} and below we denote matrices in site space with a hat and the corresponding matrix product with a star ($*$). Equation~\eqref{eq:hf_eom} is a self-consistent approximation for description of the dynamics of \(\hat{G}^<(t, t^\prime)\) in an interacting system~\cite{paper:kita_nonequilibrium}. In Ref.~\cite{paper:hf_original}, this approximation was introduced in the context of MBL. 
 \par
Observables considered in this paper are expressed in terms of the density expectation values at individual sites $j$ at time \(t\), which are related to the lesser Green's function as
\be
    \left\langle n_j (t) \right\rangle = - \mathrm{i} G^<_{j, j}(t, t)
    := - \mathrm{i} g_{j, j}(t).
\ee
Therefore it suffices for our purposes to consider the lesser Green's function with same time arguments, abbreviated by \(g_{i, j}(t)\) in the last line. According to Eq.~\eqref{eq:hf_eom}, the time evolution of \(g_{i, j}(t)\) is determined by
\begin{align}
    \mathrm{i} \partial_t \hat{g}(t) &= \left( \hat{H}_0 - \hat{\Sigma}^\mathrm{HF}(t) \right) * \hat{g}(t) - \hat{g}(t) *\left( \hat{H}_0 - \hat{\Sigma}^\mathrm{HF}(t) \right) \nonumber \\
    &:= \left[ \hat{H}_0 - \hat{\Sigma}^\mathrm{HF}(t) , \hat{g}(t)\right]. \label{eq:hf_eom_same_time}
\end{align}
We integrate this equation for successive time points by using an interface to ODEpack \cite{book:odepack} solvers contained in SciPy \cite{paper:scipy}.
\par
In the process of solving Eq.~\eqref{eq:hf_eom_same_time}, we have to compute the commutator on its right hand-side at every time step.  For the considered models with short-range hopping and interaction, the number of entries in the hopping matrix \(\hat{J}\) and the interaction matrix \(\hat{V}\) scale linearly with the number of sites \(N\). Thus, the necessary matrix multiplications can be performed with \(\mathcal{O}(N^2)\) operations. For the considered system sizes, the complexity of the solving process depends on the size mainly through the evaluation of Eq.~\eqref{eq:hf_eom_same_time}. Therefore, the computation time scales approximately as \(\mathcal{O}(N^2 \cdot N_\mathrm{time})\), where \(N_\mathrm{time}\) is the number of time steps. 
\par For the solving process, we generally split a unit physical time into \(10^2\) steps. We have checked that the chosen time step is sufficiently small to ensure that observables under consideration depend only weakly on it; see \ref{app:integration_step_size} for details.
\par
For the purpose of benchmarking, we will compare the TDHF to two different established methods. Exact results for the time evolution of an arbitrary initial state can be obtained by directly applying the time evolution operator and calculating matrix elements. The run time and memory consumption of this process consequently scale exponentially with the system size \(N\). Using the QuSpin package ~\cite{paper:quspin_ref1, paper:quspin_ref2} we can simulate up to $N = 22$ using modest numerical resources. Furthermore, we compare  to results from Refs.~\cite{paper:mbl_elmer, Doggen2019a} obtained using the TDVP with matrix product states \cite{paper:TDVP_2}.
%

\subsection{Observables}

For 1D systems, a central observable to be studied is the real-space imbalance as a function of time, which is defined as
\bea
 &&   I(t) = \frac{\left\langle n_\mathrm{even}(t)\right\rangle - \left\langle \label{eq:imbalance} n_\mathrm{odd}(t)\right\rangle}{N} \,. \\
  &&   n_\mathrm{even} = \sum_{i \ \mathrm{even}} n_i \,, \qquad n_\mathrm{odd} = \sum_{i \ \mathrm{odd}} n_i  \,.
\eea
Angular brackets here and in the following are understood as denoting averaging over states as well as over disorder configurations. 
In the quasi-periodic case, the disorder averaging is replaced by averaging over the phase shift $\phi_0$, Eq.~\eqref{phi-0}. 
Initially, we prepare the system in a staggered state where all even sites are occupied and all odd states are empty, corresponding to the maximal possible imbalance, \(I(t_0)=\unit[1]{}\).
\par
We also investigate the imbalance in energy-space, the columnar imbalance for 2D systems, as well as the first moment of the number density describing melting of a domain wall. All these observables are defined in terms of the components of \(\hat{g}(t)\) and will be introduced in respective sections of the paper below.
Errors are estimated using a bootstrapping procedure.

\paragraph{Decay of the imbalance}

Previous work suggested that the decay of the imbalance in disordered 1D systems is of power-law character
\cite{paper:mbl_powerlaws},
\begin{align}
    I(t) \sim t^{-\beta}.
\end{align}
To characterize the decay found in the simulations, it is convenient to define a flowing (time-dependent) exponent  $\beta(t)$ via
\begin{align}
    \beta(t) &= - \partial_{\log(t)} \log[I(t)].
    \label{eq:imbalance_decay}
\end{align}
We evaluate Eq.~\eqref{eq:imbalance_decay} numerically by determining the slope from a window of intermediate size that permits to average out fast fluctuations, cf.~Ref.~\cite{Doggen2019a}.


\section{One-dimensional random system}
\label{subsec:1d_random}

We consider the dynamics of a 1D randomly disordered system as described by Hamiltonian~\eqref{eq:h_fermihubbard}, where we set set \(J=\unit[0.5]{}\) and \(U=\unit[0.5]{}\).
The choice $J=U$ corresponds to an isotropic Heisenberg interaction in the spin representation. 
We first compare TDHF results to exact methods and then employ the TDHF for large systems and long times \cite{paul_thesis}.

For convenience of the reader, we mention correspondence between our notations for parameters of the model and those in some of previous works. 
The hopping strength \(J\), interaction strength \(U\), and disorder strength \(W\) are defined in our paper in the same way as in Ref.~\cite{paper:hf_original}.
Further, our model with ($J=0.5$;  $U=0.5$;  $W$) is equivalent to  the model of  Refs.~\cite{paper:mbl_elmer,mps_mbl_review} with ($J=1$;  $\Delta=1$;  $W$) and to the spin Hamiltonian of Ref.~\cite{paper:mbl_ed} with disorder strength $h = W$.

\subsection{Comparison to exact approaches}
\label{sec:1d-comp-to-exact}

We compare results obtained with the TDHF to results of a brute-force exact calculation up to \(\unit[10^4]{}\) hopping times in a chain of \(L=\unit[22]{}\) sites as well as to TDVP results from Ref.~\cite{paper:mbl_elmer} in a system of \(L=\unit[100]{}\) sites up to \(\unit[10^2]{}\) hopping times.
Our choice of units corresponds to the convention from Ref.~\cite{paper:mbl_elmer}. In accordance with Ref.~\cite{paper:mbl_elmer} we use open boundary conditions for the comparison. Disorder strengths are between \(W=\unit[2]{}\) and \(W=\unit[8]{}\).

\subsubsection{Small systems}
\begin{figure}[H]
	\centering
	\includegraphics[width=\linewidth]{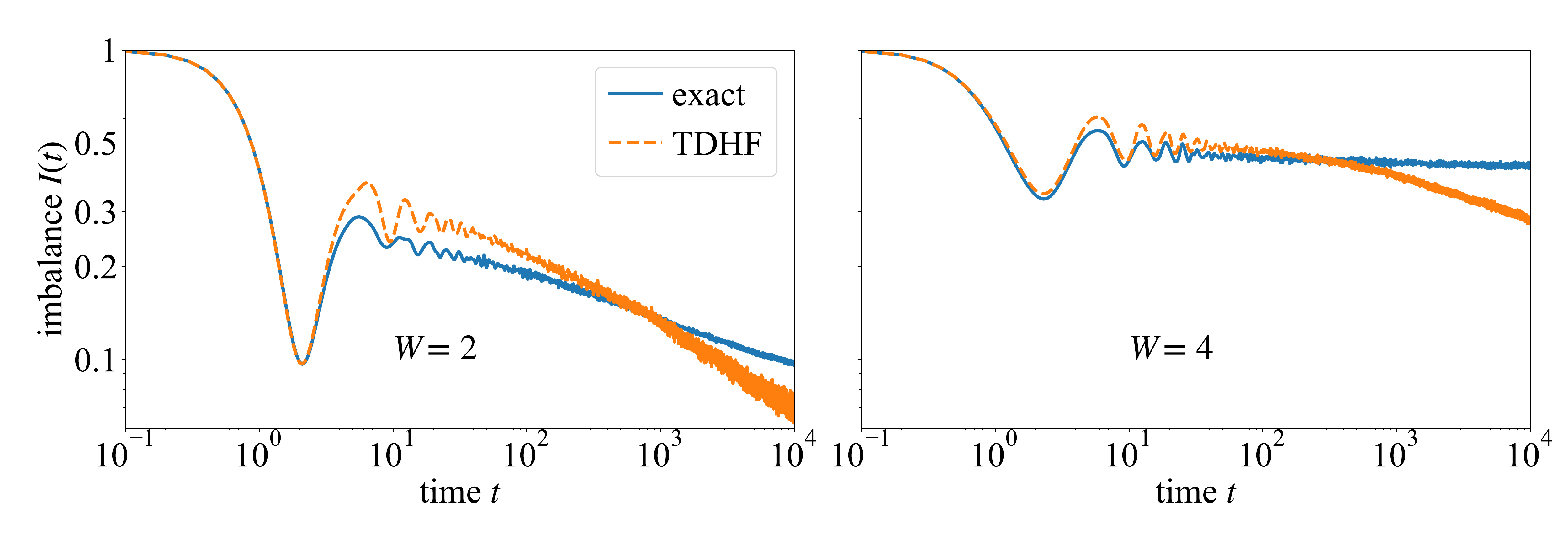}
	\caption{Imbalance as a function of time obtained from exact calculation (blue solid line) and TDHF (orange dashed line) in a 1D random system of \(L=\unit[22]{}\) sites with open boundary conditions (OBC). Exact and TDHF results were averaged over $\approx 60$ and $ \approx 200$ samples respectively. The left and right panel show the imbalance at \(W = \unit[2]{}\) and \(W=\unit[4]{}\).}
	\label{fig:imbalance_comparison_exact}
\end{figure}

\begin{figure}[H]
	\centering
    \includegraphics[width=\linewidth]{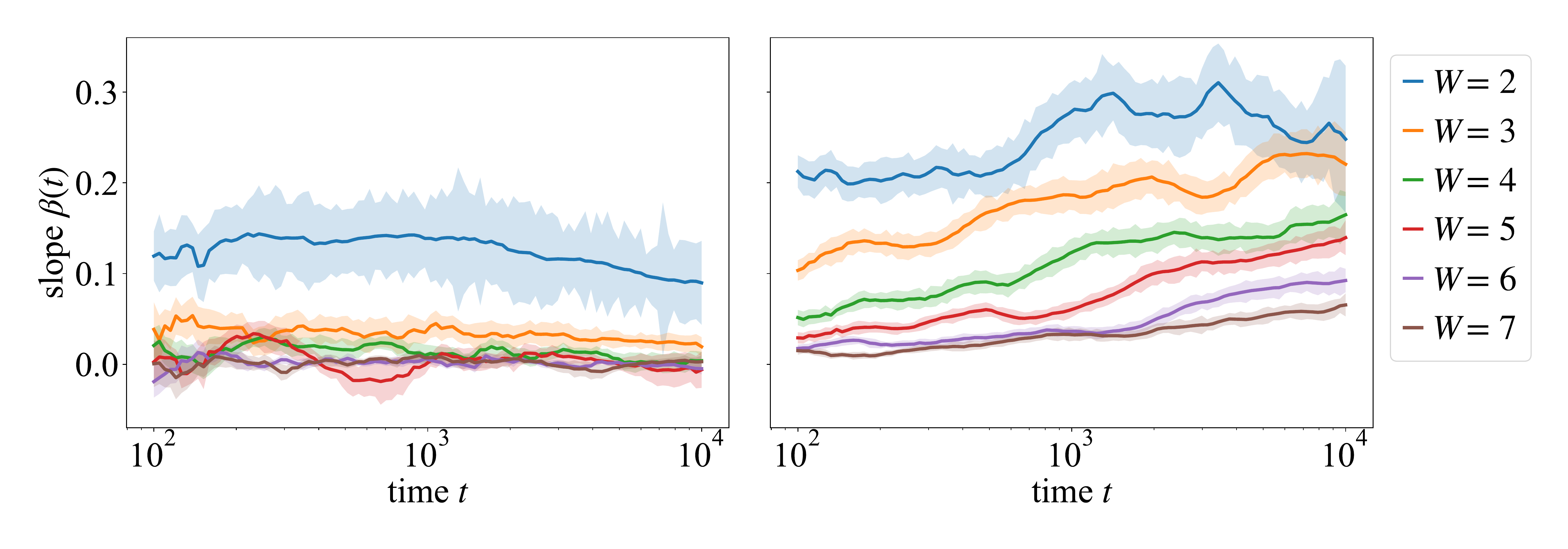}
	\caption{Time dependence of the imbalance exponent (slope) for the disorder $W=2$, 3, 4, 5, 6, and 7 in a  1D random  system of \(L=\unit[22]{}\) sites with OBC. 
	Left and right panels present exact and TDHF results, respectively.	
	The shaded regions indicate the bootstrap errors on the individual fits.}
	\label{fig:imbalance_slope_comparison_exact}
\end{figure}

First, we consider a small system with  \(L=\unit[22]{}\) sites.  Figure \ref{fig:imbalance_comparison_exact} shows the imbalance $I(t)$ as obtained from exact calculation and TDHF for $W=2$ and $W=4$. Both the exact and the TDHF curves show slow dynamics but the long-time decay within the TDHF approach is clearly faster. In particular, at \(W=4\), the exact imbalance indicates localization (for this small system size) as it remains constant above \(t \sim 10\) hopping times. In contrast, the TDHF imbalance visibly decays. Clearly, the TDHF is only an approximate method. To inspect the character of the dynamics, we show in Fig.~\ref{fig:imbalance_slope_comparison_exact} the time dependence of the slope $\beta(t)$   defined by Eq.~\eqref{eq:imbalance_decay} for the disorder interval \(W\in[2, 7]\). Exponents $\beta$ below 0.5 mean slow, subdiffusive transport.  Differences between the exact results and those of the TDHF are again manifest.  While the exact imbalance has, within statistical errors, a vanishing slope for \(W \gtrsim 4\), the TDHF imbalance shows decay up to the strongest considered disorder \(W=7\), and the corresponding exponent $\beta(t)$ increases significantly in this time range. However, we see that within the TDHF the transport remains clearly subdiffusive, with $\beta(t) < 0.3$. This is a first indication of the fact that the TDHF captures the subdiffusive character of the dynamics, even though it is obtained for rather small systems. We will see below that this remains true for large $L$ and, moreover, that the performance of the TDHF is even better in larger systems. (The latter observation is consistent with Ref.~\cite{paper:hf_original} where it was pointed out that field-theoretical approaches, like self-consistent TDHF approximation, tend to effectively reduce finite-size effects, thus mimicking larger systems.)

\subsubsection{Large systems}

\begin{figure}[H]
	\centering
	\includegraphics[width=\linewidth]{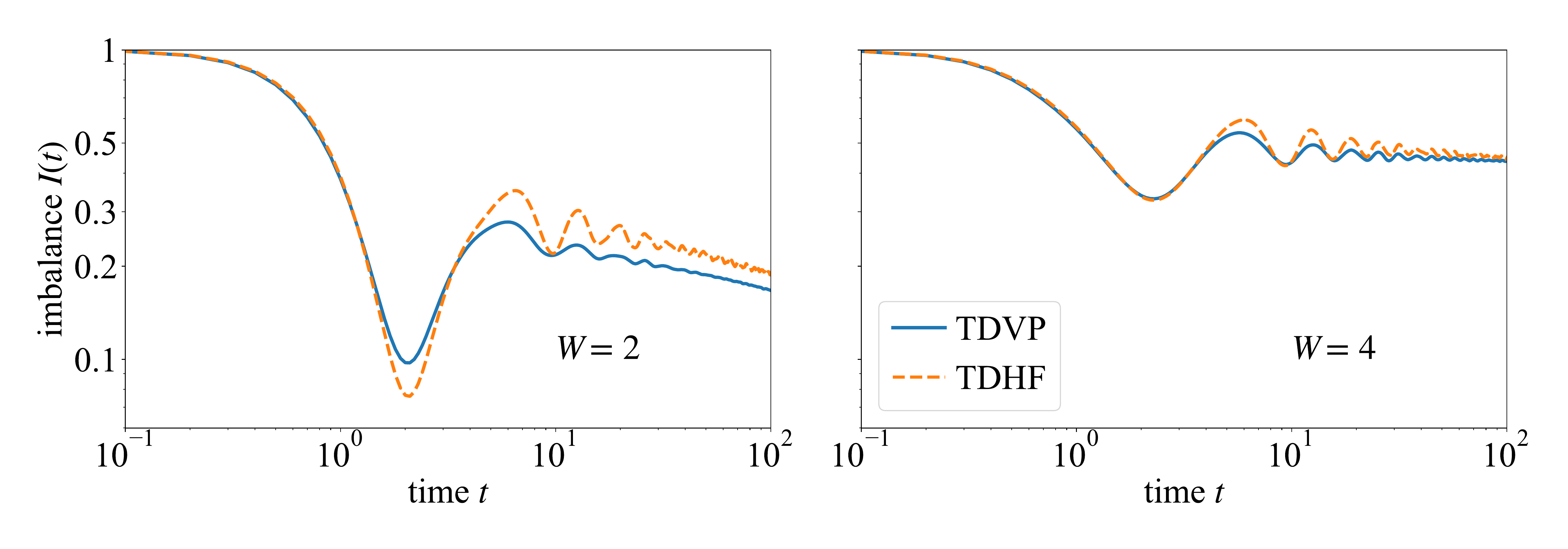}
	\includegraphics[width=\linewidth]{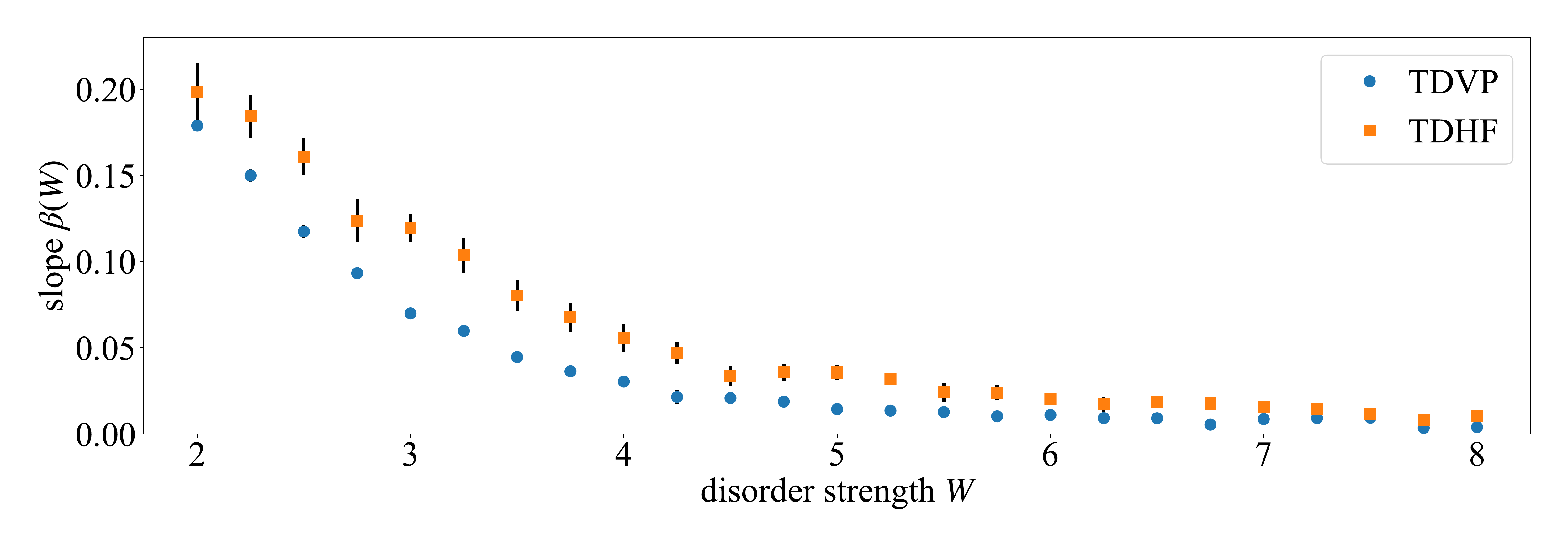}
	\caption{Comparison of imbalances as obtained from TDVP (blue) and TDHF (orange) in a  1D random  system of \(L=\unit[100]{}\) sites with OBC for disorder $W=2$ (left) and $W=4$ (right). The TDVP data is taken from Ref.~\cite{paper:mbl_elmer}. TDVP and TDHF results are averaged over  $\sim 10^3$ and $\sim 10^2$ disorder realizations, respectively. In the bottom figure, exponents $\beta$ obtained from fits in the time interval \(t \in [5\cdot 10^1, 10^2]\) are shown. The error bars (one sigma) are obtained by a bootstrapping procedure.}
	\label{fig:comparison_tdvp}
\end{figure}

We proceed now to much larger systems, further comparing the TDHF to established methods. The comparison to quasi-exact results in larger systems---i.e., those of our actual interest---is of particular importance. Clearly, brute-force numerics is not possible any more at \(L=\unit[100]{}\) sites. However, essentially exact results for not too long times, $t \lesssim 10^2$, can be obtained by using the TDVP approach with matrix product states, Ref.~\cite{paper:mbl_elmer,mps_mbl_review}.  Figure~\ref{fig:comparison_tdvp} shows a comparison between imbalances for $W=2$ and $W=4$ as obtained by TDHF and TDVP. While TDHF and TDVP imbalance values differ noticeably in both plots, the slopes are remarkably close such that a difference is barely visible. The TDHF thus indeed performs better in larger systems, as was pointed out above and in Ref.~\cite{paper:hf_original}.
To compare the slopes (i.e., the exponents $\beta$) obtained by the two methods more accurately, we plot them (calculated from data in the time window $t \in [50, 100]$) in the lower panel of Fig.~\ref{fig:comparison_tdvp} as functions of disorder in the range from $W=2$ till $W=8$. We see that the dependence $\beta(W)$  given by TDHF is qualitatively very similar and numerically quite close to the ``quasi-exact'' one (obtained by TDVP). At the same time, it is seen that TDHF somewhat overestimates $\beta$ (i.e., the dynamics is ``more delocalized'' within TDHF than it actually is). In particular, $\beta$ given by TDHF remains positive within error bars well above the critical disorder strength $W_c \approx 5.5$ found by TDVP in Ref.~\cite{paper:mbl_elmer,mps_mbl_review}.
\footnote{Recent exact-diagonalization studies of the 1D random model \eqref{eq:h_fermihubbard}  pointed out a substantial drift of the critical disorder with system size $L$ \cite{Suntajs2020a,Sels2020a}, $ \partial W_c /\partial L \simeq 0.1$ at $L \lesssim 20$.  It was conjectured in these works that this drift might be an indication of the linear increase of $W_c(L)$ in the asymptotic limit of large $L$ (which would be in strong contradiction to existing analytical theories). However, several works emphasized that this drift is an artefact of small systems (and is also present in exactly solvable models where $W_c(L\to\infty)$ is finite and known)~\cite{finite_size_localization,Sierant_2020,mps_mbl_review,rrg_mbl_review}. A significant drift of the critical disorder strength with system size in small systems is in full agreement with the MPS-TDVP  results~\cite{paper:mbl_elmer}, which show at the same time that it saturates in systems of \(L\sim 50\) sites at \(W_c \sim 5.5\). This is also in consistency with arguments in Ref.~\cite{Panda_2020} that systems of size $L \sim 50$ are required to appropriately assess the MBL transition. 
}

\subsection{Long-time imbalance dynamics}
\label{spatial-imbalance-1D}
\begin{figure}[H]
	\centering
	\begin{subfigure}[t]{\textwidth}
	  \centering
         \includegraphics[width=\linewidth]{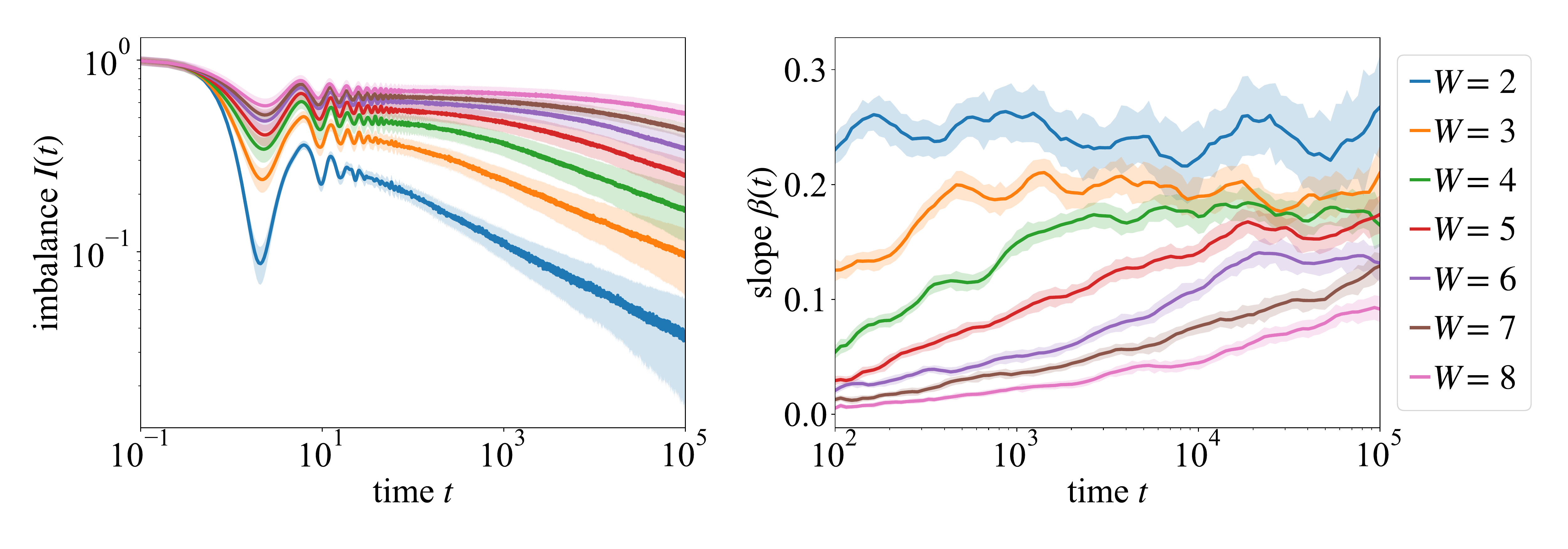}
	\end{subfigure}
	\begin{subfigure}[t]{\textwidth}
	  \centering
	  \includegraphics[width=\linewidth]{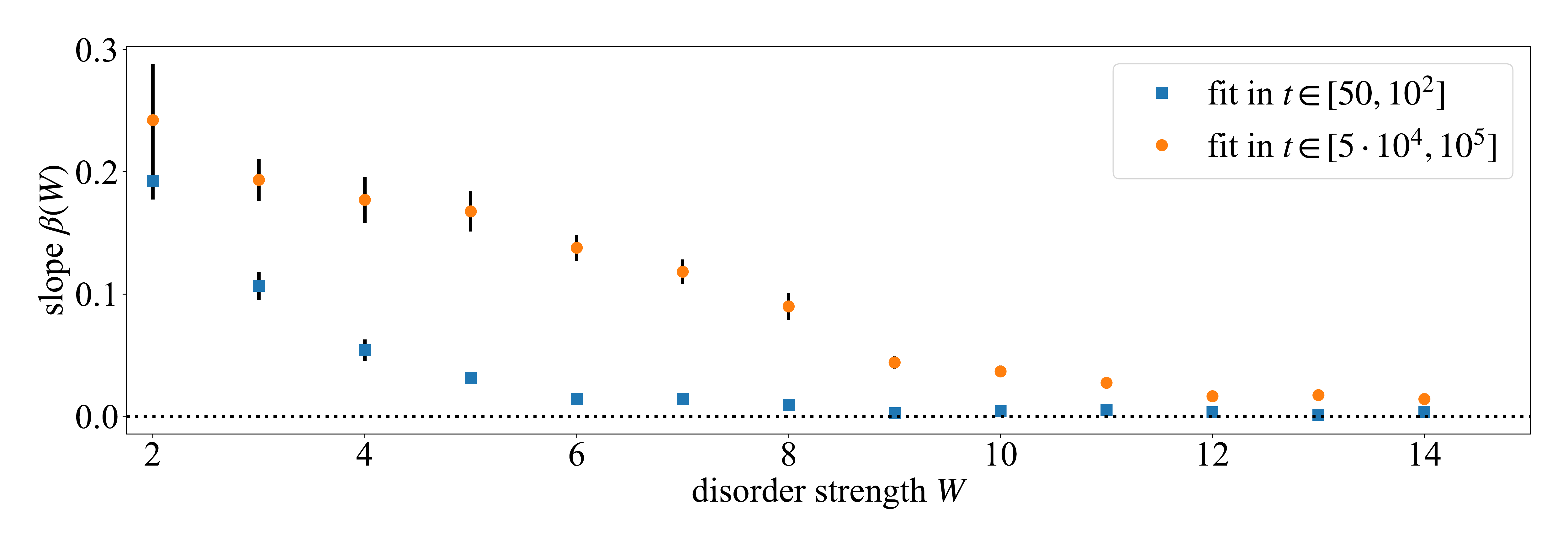}	 
	\end{subfigure}%
	\caption{Imbalance $I$ and exponent (slope) $\beta$ up to \(\unit[10^5]{}\) hopping times in a  1D random  system of \(L=\unit[100]{}\) sites at different values of disorder,
	 computed using the TDHF with OBC. The data is averaged over $ \sim 10^2$ disorder realizations. {\it Upper left:}  time dependence of the imbalance $I(t)$, 
	 for disorder strength from $W=2$ to $W=8$,
	 with the shaded regions indicating the standard deviation within the disorder sample. {\it Upper right:}  time dependence of the running imbalance exponent $\beta(t)$ (slope of the curves in left panel), with the  shaded regions indicating bootstrap errors on the individual fits. 
{\it Bottom:}	Exponents $\beta$ obtained by fits in two time windows, $[50, 100]$ and $[5 \cdot 10^4, 10^5]$,  as functions of disorder strengths in the range from $W=2$ to $W=14$,
with bootstrap errors (one sigma error bars). }
	\label{fig:long_time_hf}
\end{figure}

In Sec.~\ref{sec:1d-comp-to-exact}, we have studied the long-time behavior of small systems as well as the dynamics of large systems at relatively short times  in comparison to exact methods. We have found that, although TDHF is clearly an approximate approach, it captures one of the key properties of the problem: the subdiffusive character of the delocalized phase.
Due to the computational efficiency of the TDHF, we can extend our results for large systems (\(L=\unit[100]{}\) sites) up to much longer times. This allows us to explore, within the TDHF, the important question of whether the subdiffusive dynamics persists, in large systems, up to these long times.
The upper left panel of Fig.~\ref{fig:long_time_hf} shows the time dependent TDHF imbalance $I(t)$  up to time $t =10^5$ for disorder from $W=2$ to $W=8$.  The figures indicates that at these very long times the curves have straight-line asymptotics (on the log-log scale), which corresponds to a power-law decay of the imbalance. In order to reveal the long-time behavior of $I(t)$ in the clearest form, we plot in the upper right panel the exponent $\beta(t)$ (i.e., the slope of the curves from the left panel) in the time interval from $t=10^2$ till $10^5$.  It is seen that for the relatively weak disorder, $W=2$, the exponent $\beta(t)$ is essentially constant in the whole range of times, implying a clear power-law behavior. For stronger disorder, $W=3$, 4, 5, and 6, the running exponent first increases with time but eventually saturates. The saturation time increases with disorder strength, so that for strongest disorder in this plot ($W=7$ and 8) no saturation is reached; presumably, it requires still longer times. The lower panel of 
Fig.~\ref{fig:long_time_hf} shows $\beta(t)$ as obtained by TDHF for relatively short times ($t \in[50, 100]$; see Fig.~\ref{fig:comparison_tdvp}) and very long times ($t\in[5 \cdot 10^4, 10^5]$) in a still broader range of disorder (from $W=2$ to $W=14$). 

Two main qualitative conclusions from Fig.~\ref{fig:long_time_hf} are as follows. First, as we have just pointed out, the observed long-time saturation of $\beta(t)$ implies a power-law asymptotic behavior of the imbalance decay, $I(t) \sim t^{-\beta}$. Importantly, even for the weakest considered disorder, $W=2$, which is deeply inside the ergodic phase, the saturation value is $\beta \approx 0.25$, i.e., well below the diffusive value 0.5 \cite{note-memory}. Therefore, the TDHF approach reveals anomalous diffusion in a broad interval of disorder on the ergodic side of the MBL transition up to the very long times $t = 10^5$.  This is in agreement with the result of Ref.~\cite{paper:hf_original} (where times up to $t \sim 10^4$ were considered). 

Secondly, we do not observe a sharp MBL transition within the TDHF approximation, at variance with the conclusion of Ref.~\cite{paper:hf_original}. Specifically, even for the strongest disorder that we have considered, while the imbalance seems to stay constant at relatively short times $\sim 10^2$, it starts to drop then, with $\beta(t)$ becoming distinctly different from zero (see also~\ref{appendix:strong-disorder}). Therefore, our numerical observations indicate that the TDHF approximation destabilizes the MBL phase, producing very slow, strongly subdiffusive dynamics for those values of disorder where the dynamics should be completely frozen due to MBL. We will return to this question below and explain why this is also expected from the analytical point of view.

\subsection{Melting of a domain wall}
\label{sec:first_moment}
As a complementary probe of the dynamics of delocalization, we consider the broadening process of a domain wall initially situated in the middle of a system of \(L= \unit[50]{}\) sites. This approach is a natural way to study the character of transport; it has been used to probe the MBL physics in Refs.~\cite{Hauschild_2016_dw,paper:hf_original}. To characterize the dynamics of the melting process, we calculate the first moment of the particle density as a function of time,
\begin{align}
    x(t) = \sum_{i=1}^L i \Big[ n_i(t) - n_i(t=0) \Big].
    \label{eq:first_moment}
\end{align}
In the initial state, the sites $1 \le i  \le L/2$ have unit populations, while the other half of the system is empty, so that $x(0) = 0$.  As \(x(t)\) scales as the square of the domain-wall width, we introduce the  running exponent $\beta_{\rm dw}(t)$ characterizing the domain-wall broadening according to
\begin{align}
\beta_{\rm dw}(t) = \frac{1}{2}  \partial_{\log(t)}\log[x(t)] \,.
    \label{eq:first_moment_time_exp}
\end{align}
With this definition, one can directly compare the exponent  \(\beta_{\rm dw}(t)\) to the imbalance exponent $\beta(t)$ as both of them describe a scaling of a length scale with time~\cite{paper:mbl_powerlaws}.

\begin{figure}[H]
	\centering
    \includegraphics[width=\linewidth]{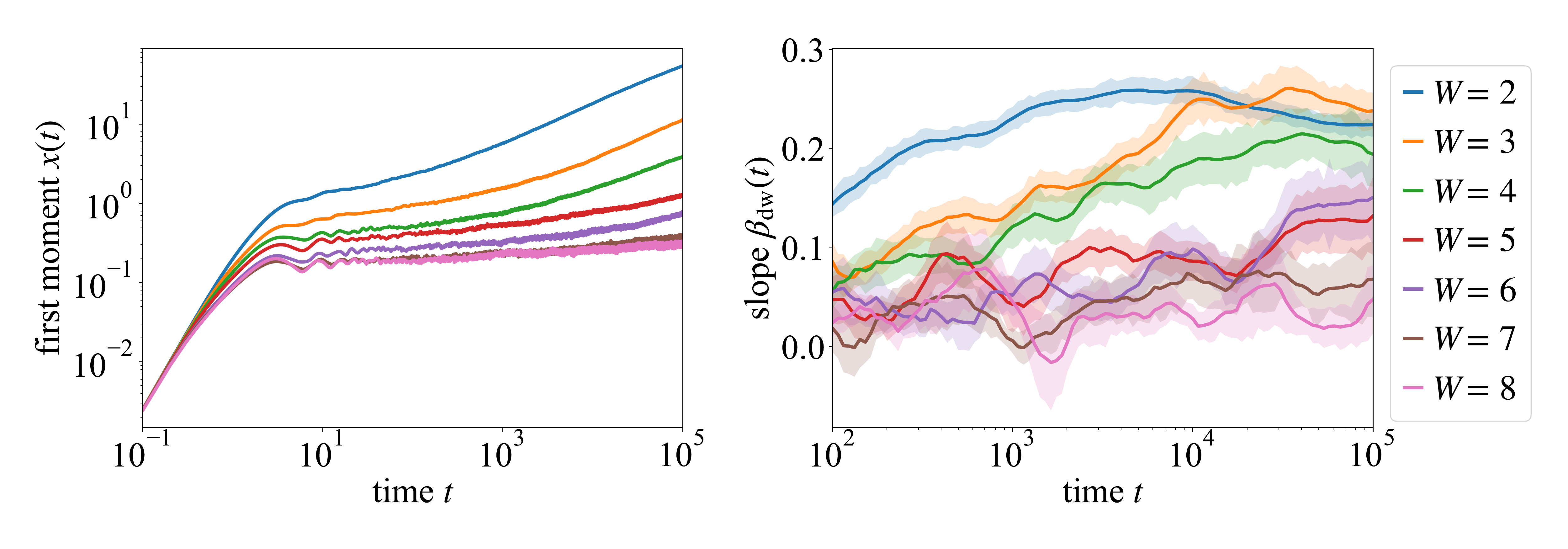}
	\caption{Melting of a domain wall in a  1D random  system of \(L=\unit[50]{}\) sites with OBC. The data is averaged over \( \sim 600\) disorder realizations. The left panel shows the first moment $x(t)$ of the particle density as a function of time, Eq.~\eqref{eq:first_moment}, and the right panel the corresponding exponent $\beta_{\rm dw}(t)$  defined by Eq.~\eqref{eq:first_moment_time_exp}.}
	\label{fig:first_moment}
\end{figure}

The results for \(x(t)\) and \(\beta_{\rm dw}(t)\) are shown in Fig.~\ref{fig:first_moment} for disorder strengths in the interval from $W=2$ to $W=8$. The system size in this plot is $L=50$, so that homogeneous occupation of the whole system would correspond to \(x(t) \rightarrow L^2 /8  \approx 300\). For our weakest disorder, $W=2$ and the longest time, $t=10^5$, the moment $x(t)$ reaches the value $\approx 50$. For other disorder strengths, $W \ge 3$ we have $x(t) \lesssim 10$  in the whole time range. We thus may expect some (relatively weak) finite-size effects at longest times for $W=2$  and no significant finite-size effects for $W \ge 3$. This is indeed what is observed: a small decrease of $\beta_{\rm dw}(t)$ for $W=2$ at $t \gtrsim 10^4$ can be presumably attributed to finite-size effects.

In general, the behavior of the domain wall exponent \(\beta_{\rm dw}(t)\) is in a good  agreement with the imbalance exponent \(\beta(t)\) from Fig.~\ref{fig:long_time_hf} although statistical fluctuations in $\beta_{\rm dw}(t)$ are stronger. The saturation values of the exponent $\beta_{\rm dw}(t)$ reached for the disorder $W=2$, 3, 4 are in the range $\approx 0.2$ --- 0.25, close to the corresponding saturation values of $\beta(t)$. The data on the domain-wall broadening thus support the conclusion on subdiffusive asymptotic behavior obtained from the analysis of the imbalance.

Imbalance dynamics and domain wall melting have also been compared (in a somewhat different way) within the TDHF in Ref.~\cite{paper:hf_original}. 
The authors of Ref.~\cite{paper:hf_original} have fitted the imbalance slope in the time window $ 10^2 \le t \le 10^4$ at disorder $W=2$ to a power-law $\sim t^{-\alpha}$ (with an exponent $\alpha$ that is analogous to our $\beta(t)$ but is time-independent). Further, they have found that the same exponent $\alpha$ characterizes the system-size scaling of the time $t^*$ at which the domain wall spreads over the whole system, $t^* \sim L^{1/\alpha}$ (for $500 \lesssim  t^* \lesssim 5000$).  Our results demonstrating the agreement between the flowing exponents $\beta_{\rm dw}(t)$ and  $\beta(t)$  in the broad disorder range and up to time $t = 10^5$ thus corroborate and reinforce the conclusion of Ref.~\cite{paper:hf_original} concerning the correspondence between these two approaches.

\subsection{Energy-space imbalance dynamics}
\label{energy-imbalance-1D}

The imbalance that was analyzed above was defined as a measure of an ``antifferromagnetic order'' in the real space. Now we analyze the anomalous dynamics by using an analogous observable defined in energy space.  To this end, we diagonalize the non-interacting part of the Hamiltonian for a given disorder realization and prepare the initial state where every second of the resulting single-particle eigenstates (sorted by energy) is occupied. Our initial state is thus an exact eigenstate of the non-interacting Hamiltonian. We time-evolve these states and calculate the imbalance between occupation of even and odd states within the above energy ordering.

\begin{figure}[H]
	\centering
	\includegraphics[width=\linewidth]{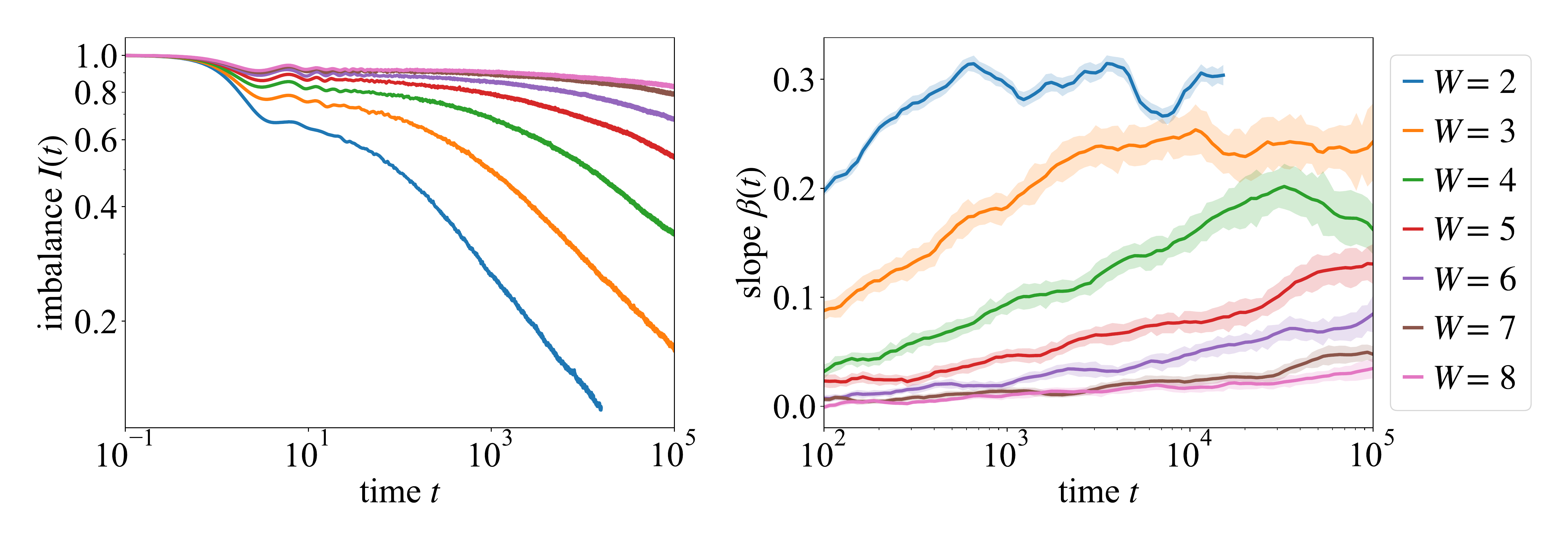}
	\caption{Energy-space imbalance dynamics in a 1D random system of \(L=\unit[100]{}\) sites with PBC, averaged over $\sim 100$ disorder realizations. The initial state is obtained by occupying every second single-particle eigenstate (in energy ordering) of the non-interacting problem. The imbalance is defined between the initially occupied and unoccupied states. Left and right panels show the time dependence of the imbalance $I(t)$ and of the exponent (slope) $\beta(t)$  [defined according to Eq.~\eqref{eq:imbalance_decay}], respectively.}
	\label{fig:energy_space_dynamics}
\end{figure}

In the left panel of Fig.~\ref{fig:energy_space_dynamics}, we show the time dependence of the energy-space imbalance for disorder from $W=2$ to $W=8$. The corresponding exponent defined via Eq.~\eqref{eq:imbalance_decay} (e.g., the slope of the curves in the left panel) is displayed in the right panel. Comparison of both panels of Fig.~\ref{fig:energy_space_dynamics} with the corresponding (i.e., upper) panels of Fig.~\ref{fig:long_time_hf} shows a clear similarity between the dynamics of the real-space and energy-space imbalance. The saturation values are rather close for both types of imbalances, which makes it plausible that the long-time $t^{-\beta}$ asymptotics for both of them is characterized by the same value of the exponent $\beta$. 

Our results demonstrate that the anomalous slow dynamics is rather generic and holds for very different types of initial conditions. In this context, we mention that it was proposed in Ref.~\cite{paper:hf_original} that an exact eigenstate of the non-interacting Hamiltonian would not experience thermalization within the TDHF approximation. Our findings indicate that this is not the case: such a state thermalizes in a way rather similar to the thermalization of an initial state with ``antiferromagnetic'' site occupation as considered in Sec.~\ref{spatial-imbalance-1D}. In fact, since each single-particle state is localized around a certain site, and since nearby-in-energy single-particle states are far away in real space, there is some qualitative similarity between our initial state with maximal energy-space imbalance and initial states with random site occupation investigated in Ref.~\cite{paper:hf_original}. Indeed, the corresponding imbalance traces $I(t)$ look quite similar. 

\section{One-dimensional quasi-periodic system}  
\label{subsec:quasiperiodic_system}

In Sec.~\ref{subsec:1d_random} we have studied the many-body delocalization dynamics in 1D random systems by self-consistent TDHF approach. The results obtained from the analysis of the real-space imbalance were supported by the investigation of the energy-space imbalance and domain-wall broadening. For all of them, we have numerically determined the running exponent $\beta(t)$. All the methods give consistently strong evidence of the slow, subdiffusive transport with $\beta < 0.3$ for the disorder range $W \ge 2$ (which includes a major part of the ergodic phase) at asymptotically long times (or, at least, up to very long times $t \sim 10^5$ studied in our work). A plausible explanation of the subdiffusive dynamics is based on Griffiths physics associated with rare events.

In order to shed more light on these results, it is important to compare them with those obtained for other types of systems, where the rare-event physics is expected to have different manifestations or not to be operative at all. This is done in the present section, where we consider 1D systems with quasi-periodic potentials, and in Sec.~\eqref{subsec:two_d}, where 2D random systems are studied. 

In this section we consider the real-space imbalance $I(t)$ in 1D chains described by the interacting Aubry-Andr\'{e} model. 
This model was defined in Sec.~\ref{sec:model}; our choice of parameters here is \(J=\unit[0.5]{}\), \(U=\unit[0.5]{}\) and \(\Phi = (\sqrt{5} - 1)/2\). The system length was $L=50$ for most of these simulations; we also studied $L=100$ systems to check the role of finite-size effects, as specified below. We first compare the TDHF results at relatively short times to those obtained by TDVP and then proceed with the analysis of dynamics at much longer times (inaccessible to TDVP). 

The interacting 1D Aubry-Andr\'{e}  model has been studied with the TDHF in Ref.~\cite{paper:hf_original}  where a faster imbalance decay that in a random model was found. Below we explore and quantify this difference by investigating the dynamics in the interacting 1D Aubry-Andr\'{e}  model up to time $t=10^5$ and determining the corresponding exponent $\beta(t)$ for various strengths $W$ of the quasi-periodic field.  Quasi-periodic 1D chains with interactions were also studied within a method similar to TDHF in Ref.~\cite{Varma_znidaric_2019} in the framework of interacting quasi-periodic Fibonacci model. Our results presented below cannot be directly compared to Ref.~\cite{Varma_znidaric_2019} since the two models are essentially different.

\subsection{Comparison to TDVP}

\begin{figure}[H]
	\centering
	\includegraphics[width=\linewidth]{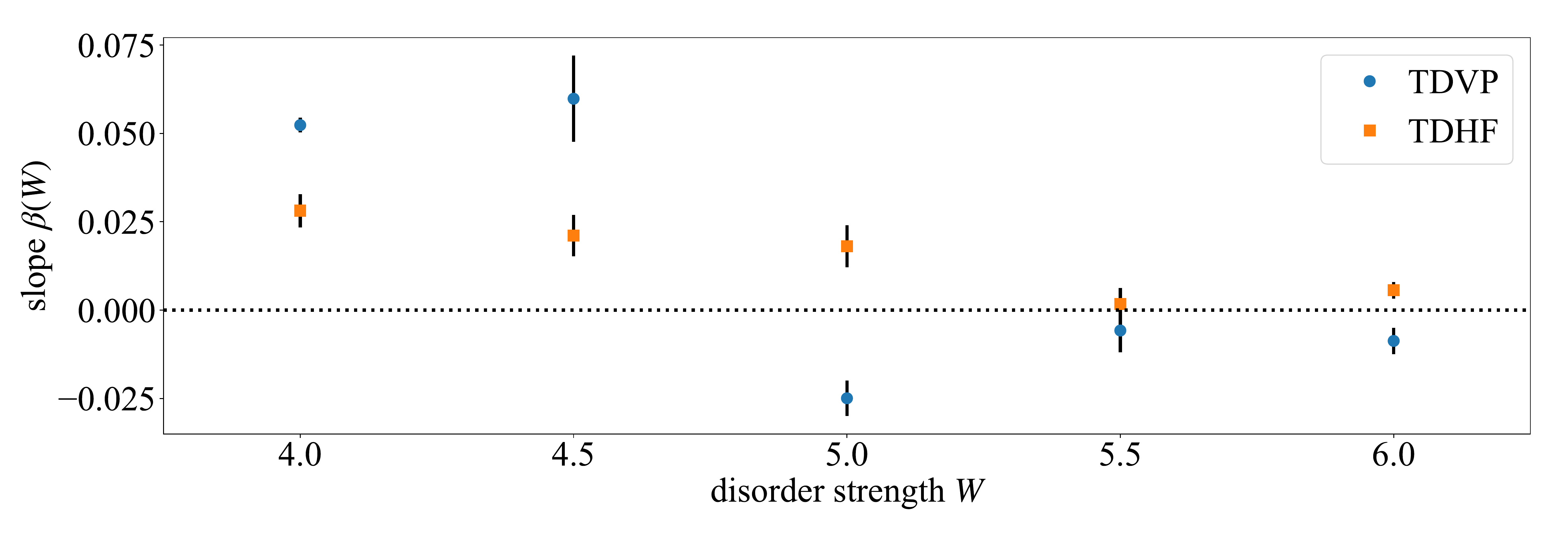}
	\caption{Comparison of imbalance exponents $\beta$ in the time interval \(t \in [50, 180]\) obtained by TDVP (from Ref.~\cite{Doggen2019a}) and TDHF in a system of \(L=\unit[50]{}\) sites with OBC. The TDVP and TDHF imbalances are averaged over $ \approx 400$ and $\approx 300$ samples, respectively. The horizontal dotted line is $\beta =0$. Error bars are two sigma, as in Ref.~\cite{Doggen2019a}.}
	\label{fig:quasiperiodic_comparison_TDVP}
\end{figure}

In Fig.~\ref{fig:quasiperiodic_comparison_TDVP}, we compare the exponent $\beta$ of the imbalance $I(t)$ obtained from TDHF simulations in the time interval \(t \in[50, 180]\) to the corresponding TDVP data of Ref.~\cite{Doggen2019a} (which are essentially exact). The disorder interval, from $W=4$ to $W=6$, includes the MBL transition point $W_c = 4.8 \pm 0.5$ as found in Ref.~\cite{Doggen2019a}.\footnote{It is interesting to note that a recent exact-diagonalization study of the stability of local integrals of motion~\cite{sarang_lioms_quasiperiodic} yielded the value $W_c = 4.0$--$4.5$ for the transition point in this model, which is close to the TDVP result $W_c \approx 4.8$ of Ref.~\cite{Doggen2019a} obtained for large systems. This is consistent with the understanding that finite-size effects in $W_c$ are relatively weak for quasiperiodic models.} 
We see that the behavior of the TDHF exponent compares reasonably well with the TDVP one. It should be noted that, for a quasi-periodic system, the exact imbalance shows oscillations on a time scale of order $t \sim 100$, see Ref.~\cite{Doggen2019a}, which leads also to oscillations of $\beta$ when determined on such time scales. In particular, as a result of such oscillations, $\beta$ provided by TDVP increases a little from $W=4$ to $W=4.5$ and then drops to a negative value at $W=5$. The TDHF approximation somewhat smears these oscillations.

\begin{figure}[H]
	\centering
    \includegraphics[width=\linewidth]{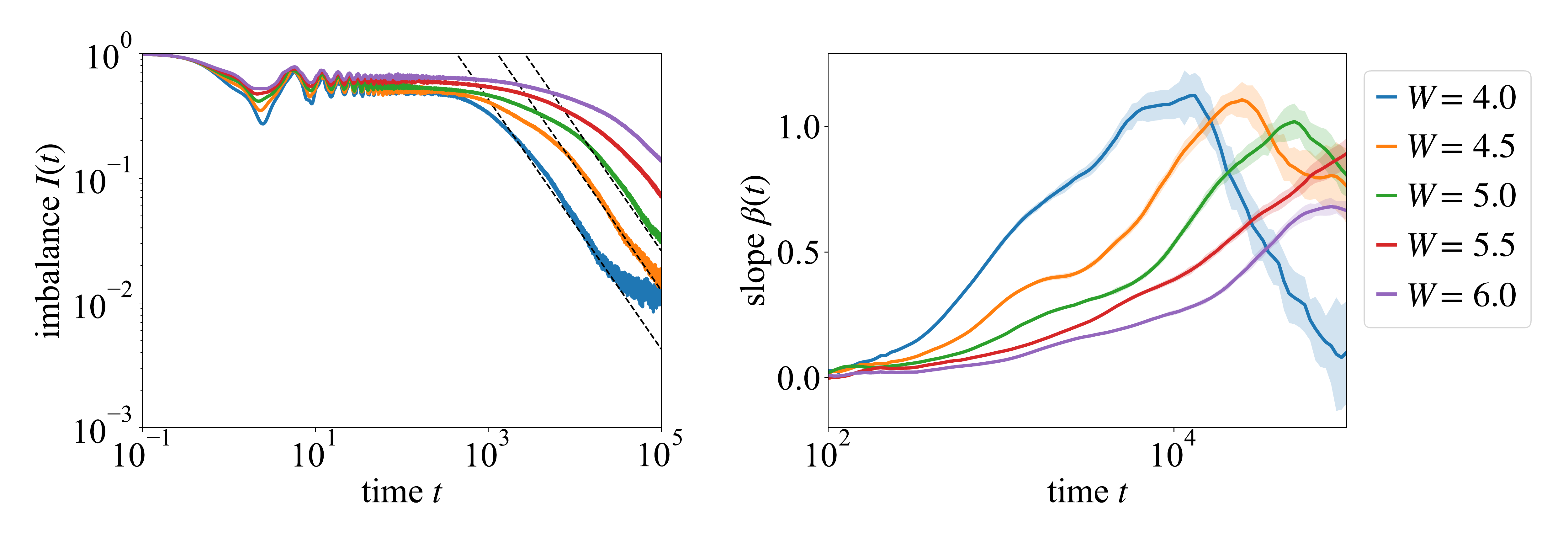}
    \includegraphics[width=\linewidth]{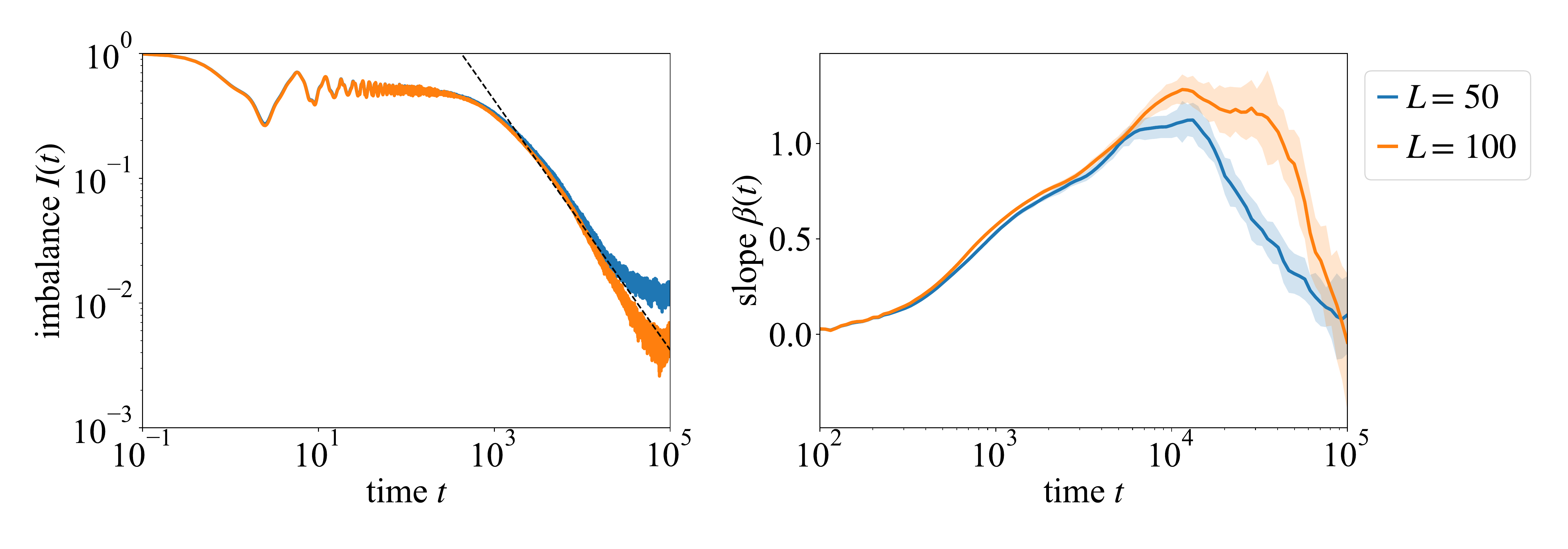}
	\caption{Imbalance dynamics in a quasi-periodic system with OBC. The data is averaged over $\approx 300$ disorder realizations. {\it Top panels:}  Imbalance $I(t)$ (left) and the running exponent $\beta(t)$ (right) for \(L=\unit[50]{}\)  and different strengths of disorder from $W=4$ to $W=6$. {\it Bottom panels:}  Imbalance $I(t)$ (left) and the running exponent $\beta(t)$ (right)  at disorder strength \(W=\unit[4]{}\) for two system lengths:  \(L=\unit[50]{}\) and \(L = \unit[100]{}\). Dashed lines in the left panels indicate the slope \(\beta = 1\) as a guide for the eye.}
	\label{fig:quasiperiodic_tail}
\end{figure}

\subsection{Long-time imbalance dynamics}

In analogy with the model with true randomness, we extend the TDHF analysis up to \(t = \unit[10^5]{}\) hopping times. The results for disorder between $W=4$ and $W=6$ are shown in  the top panels of Fig.~\ref{fig:quasiperiodic_tail}: the left panel displays the  imbalance $I(t)$ and the right panel the corresponding running exponent (slope) $\beta(t)$. 
As for the random system, we observe an initial increase of the slope with time. However, there is a crucial difference: for the quasi-periodic disorder $\beta(t)$ increases much faster and saturates at a value $\beta \approx 1$. After this saturation, $\beta(t)$ decays again down to zero. This decay is related to finite size of the system. To demonstrate this, we compare in the lower panels of Fig.~\ref{fig:quasiperiodic_tail} the data for the same disorder strength, $W=4$, and two different system sizes, $L=50$ and $L=100$. We see that an increase of the system size indeed leads to a broader plateau at $\beta \approx 1$, so that the decay of $\beta(t)$ is shifted to still later times. This confirms the plateau at $\beta =1$ in the limit of large $L$ and the finite-size character of the decay of $\beta(t)$ after this plateau. These results for quasi-periodic systems confirm that the slow, subdiffusive transport (with $\beta < 0.3$)  found for random systems in Sec.~\ref{subsec:1d_random} is indeed due to Griffiths effects.

\begin{figure}[H]
	\centering
	\includegraphics[width=0.8\linewidth]{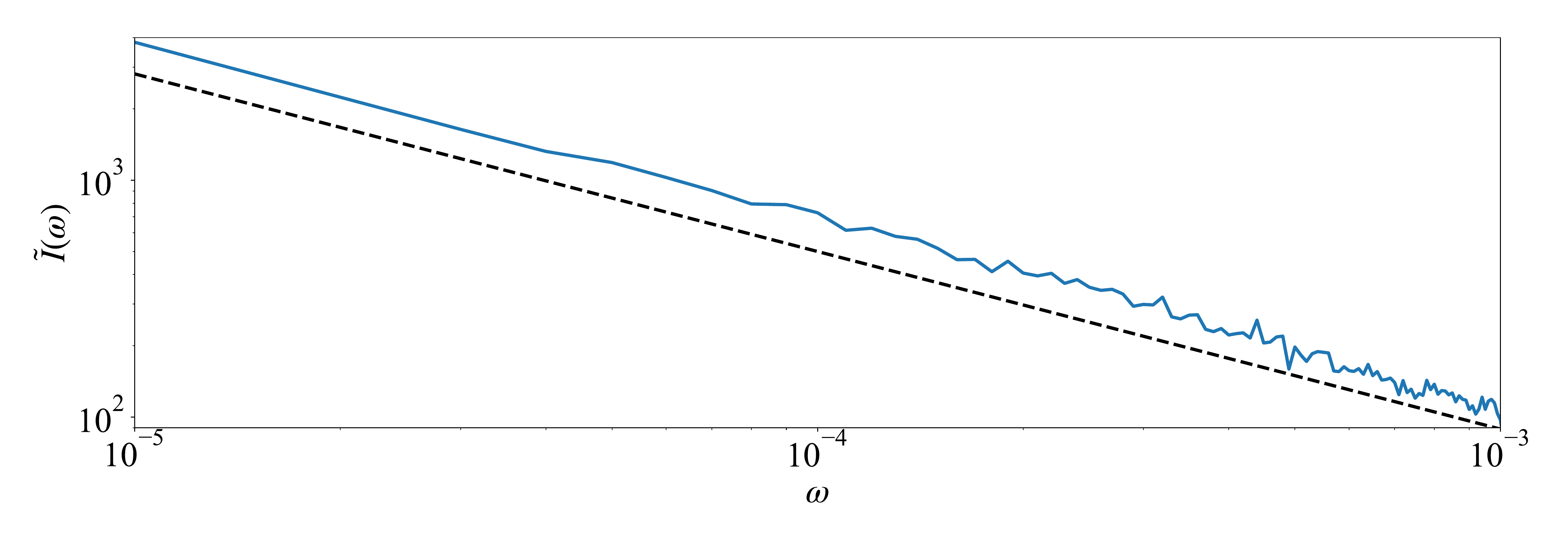}
	\includegraphics[width=0.8\linewidth]{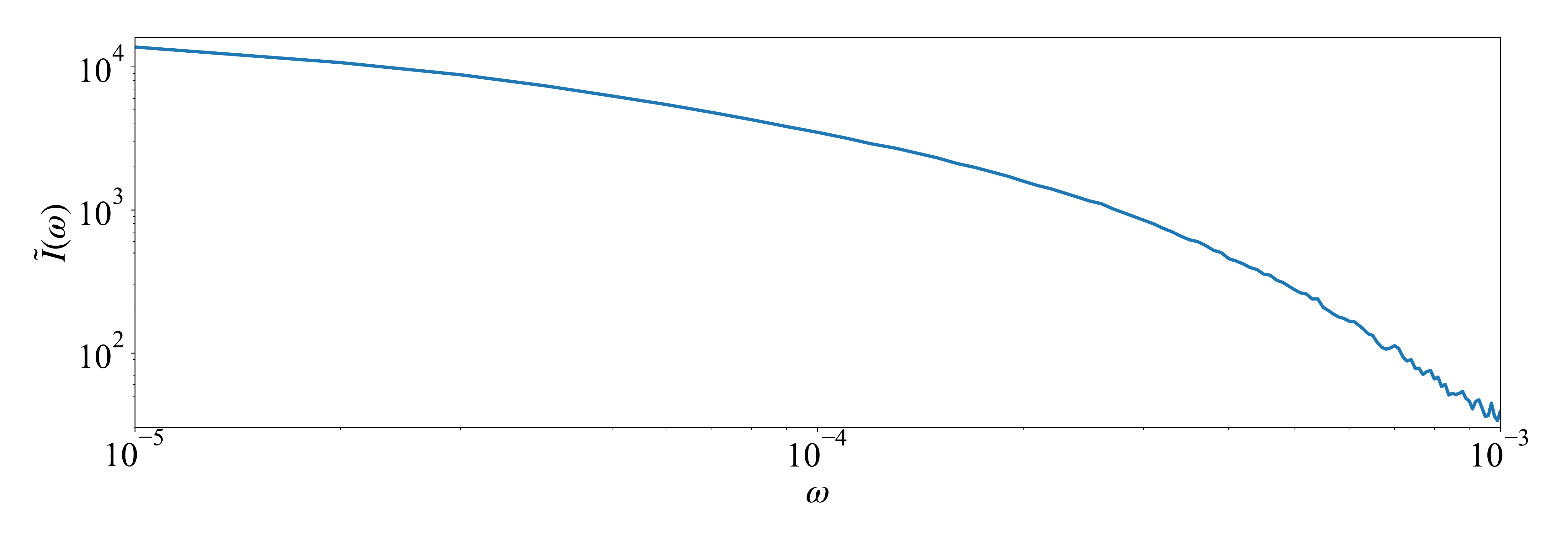}
	\caption{The real part of the Fourier transform $\tilde{I}(\omega)$ of the imbalance $I(t)$  for random (upper panel; $W=2$) and quasi-periodic (lower panel; $W=4$) 1D systems of size $L=100$.  For the random system a straight-line fit with a slope $-0.75$ [i.e., $\tilde{I}(\omega) \sim \omega^{-0.75}$] is shown.}
	\label{fig:I-omega-random-quasiper}
\end{figure}

The strong difference between the dynamics in random and quasi-periodic 1D systems is also illustrated in Fig.~\ref{fig:I-omega-random-quasiper} where we show the real part of the Fourier transform $\tilde{I}(\omega)$ of the imbalance $I(t)$.  The presented data correspond to the weakest disorder strength that we have studied: $W=2$ for random system and $W=4$ for quasi-periodic system. For the random system (shown in the upper panel), $\tilde{I}(\omega)$ is fitted very well by a straight line on the log-log scale, which means a power-law dependence $\tilde{I}(\omega) \sim \omega^{\beta-1}$, with $\beta \approx 0.25$. This value of $\beta$ is in full correspondence with the one obtained from the analysis of the running exponent in time representation, see upper right panel of Fig.~\ref{fig:long_time_hf}.   
At the same time, for the quasi-periodic case, the $\tilde{I}(\omega)$ dependence shows a strong curvature on the log-log scale and becomes nearly flat for the smallest $\omega$, in correspondence with the $1/t$ long-time behavior of $I(t)$.

The saturation of the exponent at $\beta =1$ for quasi-periodic systems implies ballistic long-time dynamics. This value can be expected, for following reasons. It is well known that the transport in delocalized phase of the non-interacting Aubry-Andr\'{e} model is ballistic, so that it is natural to expect the same character of transport in the many-body delocalized phase of an interacting Aubry-Andr\'{e} system. Of course, the imbalance probes density relaxation not at small wave vector $q$ but rather at $q=\pi/a$, where $a$ is the lattice constant. However, in analogy with the case of random systems (see the comment \cite{note-memory}), we expect that mode coupling will induce a $1/t$ decay of the imbalance.

\section{Two-dimensional random system}
\label{subsec:two_d}

In this section, we investigate the long-time delocalization dynamics in 2D systems determined by the Hamiltonian~\eqref{eq:h_fermihubbard} with random potential on square lattices. We set the parameters \(J=\unit[1]{}\) and \(U=\unit[1]{}\) and consider different values of disorder strength $W$. The boundary conditions are chosen periodic in one direction and open in the other direction, as  in Ref.~\cite{paper:elmer_2d}, so that the system can be viewed as surface of a cylinder.  Following Ref.~\cite{paper:elmer_2d}, we use as an observable the ``columnar imbalance'' between rings of sites. In the initial state, every second ring is occupied, so that the columnar imbalance has its maximal value (unity).

\begin{figure}[H]
	\centering
	\includegraphics[width=\linewidth]{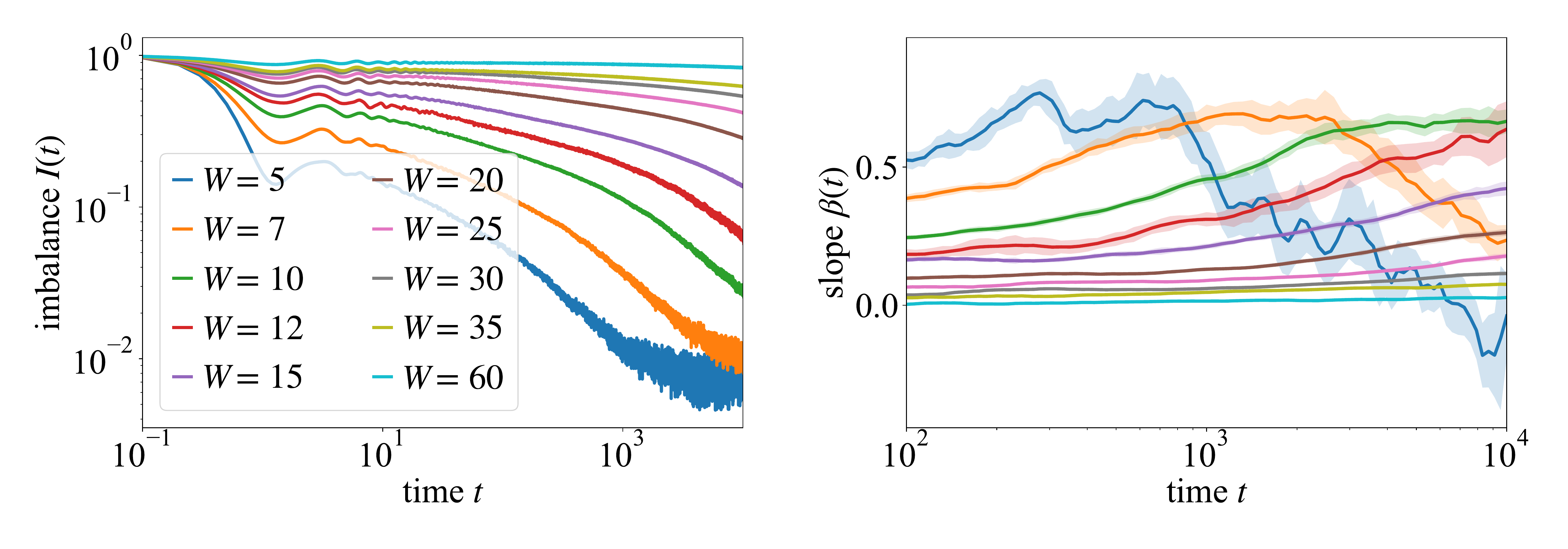}
    \includegraphics[width=\linewidth]{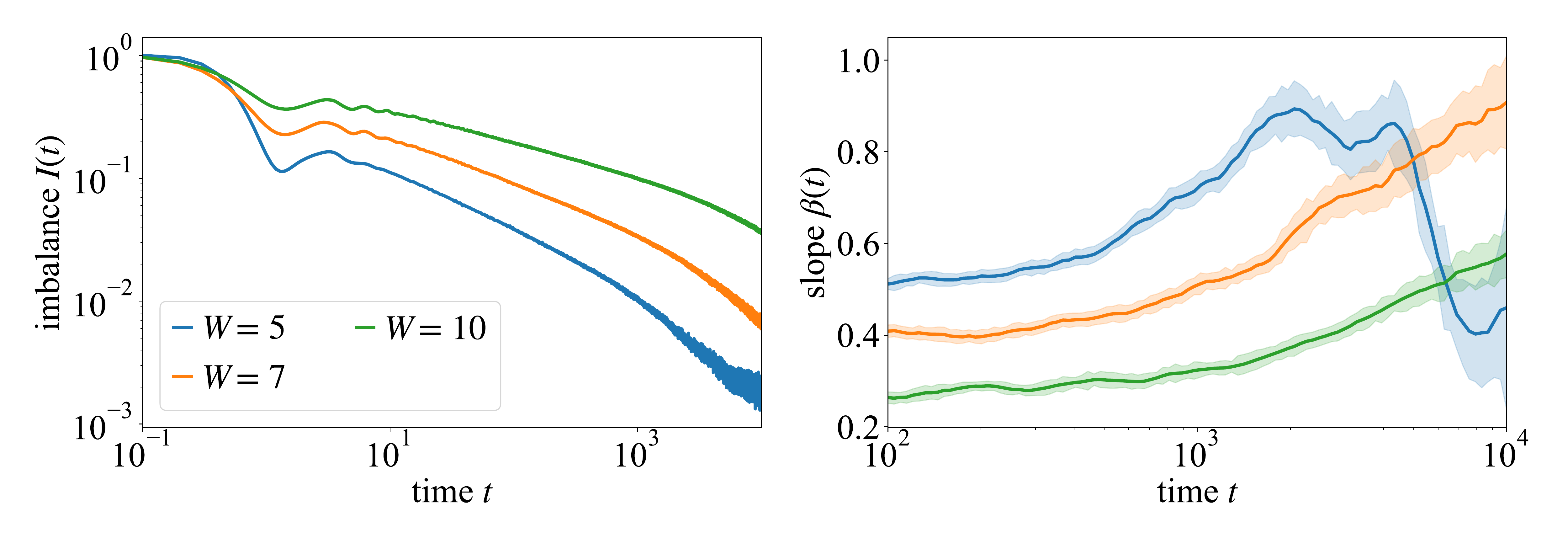}
    \caption{Dynamics of columnar imbalance in a 2D random system. The boundary conditions are periodic along one direction and open along the other. {\it Upper panels:}  system size \(10 \times 10\) sites; disorder from $W=5$ to $W=60$.  {\it Lower panels:} system size \(20 \times 20\) sites; disorder $W=5$, $W=7$, and $W=10$.    Left panels show the imbalance $I(t)$, and right panels the corresponding exponent $\beta(t)$. Averaging is performed over $\approx 300$ samples. }
	\label{fig:2d_dynamics}
\end{figure}

The obtained results for the columnar imbalance $I(t)$ and for the corresponding running exponent $\beta(t)$ are shown in upper panels of Fig.~\ref{fig:2d_dynamics} for systems of \(10 \times 10\) sites in the disorder range from $W=5$ to $W=60$. The exponent $\beta(t)$ increases with time, reaching values considerably larger than in the 1D random case. In particular, for disorder strengths $W=5$ and $W=7$ the maximal values are $\beta \approx 0.8$ and $\beta \approx 0.7$, respectively.  After reaching these maximal values, the exponent $\beta(t)$ again decreases. As we have already discussed in 
Sec.~\ref{subsec:quasiperiodic_system}, this decay is an evidence of finite-size effects. To explicitly demonstrate this, we show in the low panels of Fig.~\ref{fig:2d_dynamics} the data for disorder 
$W=5$, $W=7$, and $W=10$ in larger systems---of size \(20 \times 20\) sites. We see that now $\beta(t)$ keeps increasing towards $\beta =1$. We argue that in the limit of large $L$ and long $t$ there will be an extended plateau at the 2D diffusive value $\beta=1$ due to mode coupling (see the corresponding discussion in Sec.~\ref{subsec:quasiperiodic_system} and the comment \cite{note-memory}).

\begin{figure}[H]
	\centering
	\includegraphics[width=\linewidth]{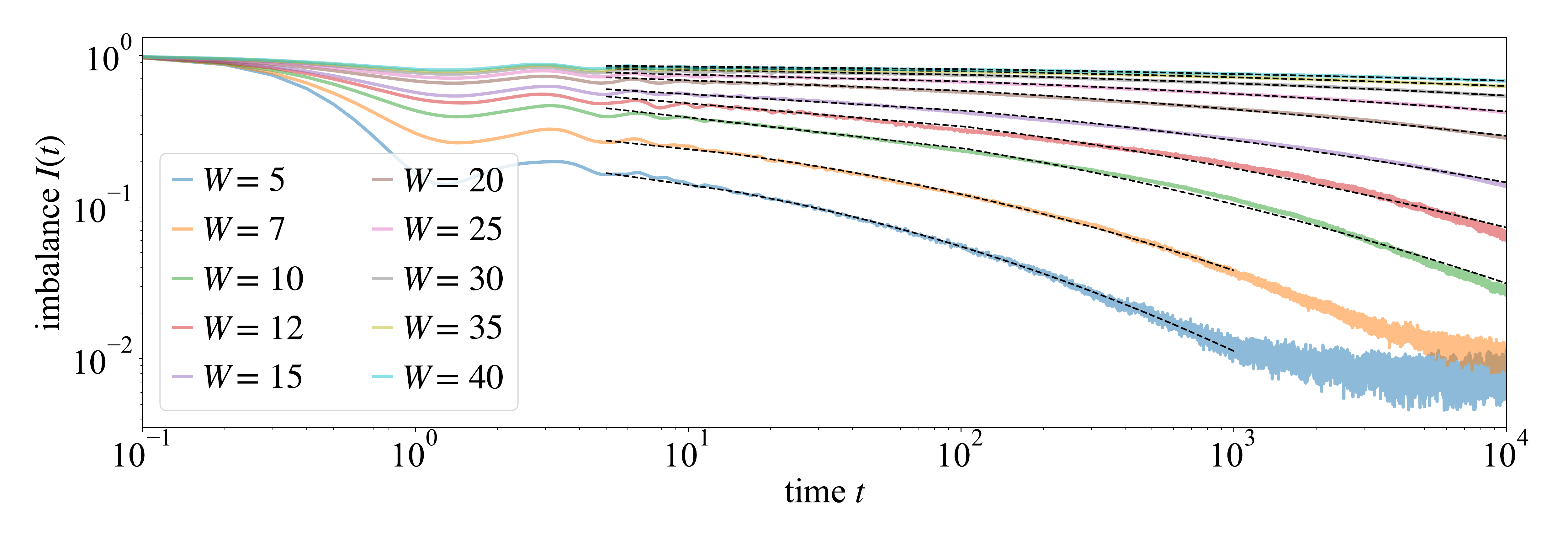}
	\caption{Fitted columnar imbalance in a 2D random system of \(10 \times 10\) sites (the data from the upper left panel of Fig.~\ref{fig:2d_dynamics}).  The figure shows the columnar imbalance (solid, in color) and fits to Eq.~\eqref{eq:fit_func_2d} (dashed, black). All imbalance curves are fitted in the time interval \([5, 10^4]\), with the exception of $W=5$ and $W=7$ fitted in the time interval \([5, 10^3]\) to exclude strong finite-size effects at \(t\gtrsim 10^3\). 
	}
	\label{fig:2d_logfit_vs_linfit}
\end{figure}

While the asymptotics at longest time in the 2D case is $\beta = 1$, one can ask a question about an increase of $\beta(t)$ before this asymptotics is reached. Assuming that this increase is determined by rare events, i.e., by generalizing the arguments that lead to a power-law behavior in the 1D case, one finds  \cite{Gopalakrishnan2019a,paper:elmer_2d} 
\begin{align}
	f(t) \sim t^{-\gamma \ln t} = e^{- \gamma \ln^2 t} \,.
	\label{eq:fit_func_2d}
\end{align}
for the columnar imbalance in two dimensional systems. As shown in Fig.~\ref{fig:2d_logfit_vs_linfit}, our data for $\beta(t)$ are fitted very well by Eq.~\eqref{eq:fit_func_2d}. 


\section{TDHF and MBL: analytical considerations}
\label{sec:analytical}

In our numerical analysis of 1D random systems in Sec.~\ref{subsec:1d_random}, we have pointed out that, within the TDHF approximation, delocalization takes place at any disorder strength. This is seen, e.g., in Fig.~\ref{fig:long_time_hf}: the exponent $\beta(t)$ increases with time up to the longest times for disorder strengths that are well above the actual MBL transition. This means that the TDHF approximation produces finite dephasing, leading to finite quasiparticle life time, relaxation, and therefore delocalization, even for strong disorder, where exact solution would yield MBL (i.e., no dephasing and relaxation). Therefore, there should be delicate cancellations in the exact equations of motion for the Green's functions that are not fully respected by the TDHF approximation. In this section, we provide the corresponding analytical considerations based on an expansion of the equations of motion in powers of interaction. First we show that, within the self-consistent Hartree-Fock approximation, such dephasing terms indeed emerge in the second order in interaction $V$. Second, we shed light on the mechanism of cancelation between the Hartree-Fock terms and those arising from the second-order self-energy in the exact solution in MBL phase. Technical details of the analysis in this section are relegated to \ref{sec:qp_broadening}.

In the TDHF approximation, the self-energy $\Sigma$ depends linearly on the Green's function:
\begin{align}
\Sigma_{\alpha\beta}^{\rm HF}(t) = V \mathcal{M}^{\alpha\beta}_{\gamma\delta} G_{\gamma\delta}(t,t) \label{eq:hartree_self}.
\end{align}
Greek indices here and in the following are used to denote components in the eigenbasis of the noninteracting Hamiltonian, distinguishing them from position-basis components for which we use latin indices as in Sec.~\ref{sec:methods}.
We separate the interaction strength $V$ from the  remaining part of the interaction matrix element $\mathcal{M}^{\alpha\beta}_{\gamma\delta}$ in order to
make the perturbative expansion in powers of $V$ more clear:
\begin{align}
G &= G^{(0)} + V G^{(1)} + V^2 G^{(2)} + \mathcal{O}(V^3),
& \Sigma&= V \Sigma^{(1)} + V^2 \Sigma^{(2)} + \mathcal{O}(V^3).
\label{eq:perturbative_scheme}
\end{align}
We tackle the self-consistent equation by solving for $G^{(i)}$ and then computing $\Sigma^{(i+1)}$ using Eq.\eqref{eq:hartree_self} order by order in $V$,  with the time evolution of all expressions starting at $t = 0$. We refer the reader to \ref{sec:qp_broadening} for details of this perturbative solution scheme. 

%

Having obtained the Green's function from Eq.~\eqref{eq:perturbative_scheme}, we then compute its Wigner-transform 
$\tilde{G}_{\alpha\alpha}^{<}(t,\epsilon)$ up to second order in $V$. Most of resulting terms only renormalize the quasiparticle weight or energy. 
However, there is also a second-order contribution [to be denoted $\tilde{G}_{\alpha\alpha}^{<,(2)b}(t,\epsilon)$] that introduces broadening of the level $\omega_{\alpha}$:
\begin{align}
\tilde{G}^{<,(2)b}_{\alpha\alpha}(t,\epsilon)
&=2\pi \mathrm{i} \sum_{\gamma\mu\nu}\mathcal{M}^{\alpha\gamma}_{\nu\nu} \mathcal{M}^{\alpha\gamma}_{\mu\mu}
\dfrac{\rho_\mu\rho_\nu}{(\omega_{\alpha}-\omega_{\gamma})^2}
\nonumber
\\
&\times
\left[
2\cos[(\omega_{\alpha}-\omega_{\gamma})t]
\delta\left(\epsilon-\frac{\omega_\alpha+\omega_\gamma}{2}\right)
(\rho_\gamma-\rho_{\alpha} )
-\delta(\epsilon-\omega_\gamma)
\rho_{\gamma} 
\right],
\label{eq:broadening_hf}
\end{align}
where  $\rho_{\alpha}$ are initial populations of levels $\alpha$.
This shows that the $\delta$-functions $\delta(\epsilon-\omega_{\alpha})$ corresponding to the non-interacting levels $\omega_{\alpha}$ are broadened. This is the source of the Hartree-Fock many-body delocalization observed in the simulations. Since the interaction is short-ranged, an essential contribution to the sum in Eq.~\eqref{eq:broadening_hf} is given by states $\gamma,\mu,\nu$ that are located close (within the distance of order of single-particle localization length) to the state $\alpha$. 

At weak interaction, the 1D random system is in the MBL phase, so that there should be no dephasing if the problem is solved exactly. Thus, the TDHF broadening \eqref{eq:broadening_hf} should be compensated by another contribution. Since such a  compensation is only possible between terms of the same order in interaction $V$, we should inspect other terms (not included in TDHF approximation) in the Green function that are of the second order in $V$. 

To this end, we include the second order self-energy (discarded by TDHF approximation)
\begin{align}
\Sigma_{ij}^{(D_2),<}(t,t')&=\sum_{k,l} V_{il} V_{jk} G_{kl}^>(t',t)\left[G_{lk}^<(t,t')G_{ij}^<(t,t')-G_{lj}^<(t,t')G_{ik}^<(t,t')\right] \label{eq:self2}
\end{align}
into the perturbative scheme where we used standard notations, with the superscripts  $>$ and $<$ denoting greater and lesser Green's functions. 
Substituting the second in the equation of motion for the Green's function, we find the 
following second-order contribution to the Green's function (see \ref{sec:qp_broadening} for detail):
\begin{align}
&\tilde{G}_{\alpha\alpha}^{<,(2')b}(t,\epsilon)=4\pi \mathrm{i}\sum_{\mu\nu\gamma}
\dfrac{\mathcal{M}^{\nu\alpha}_{\mu\gamma} \mathcal{M}^{\alpha\mu}_{\gamma\nu}}
{(\omega_\alpha -\omega_\gamma +\omega_\nu +\omega_\mu )^2}
\left\{(\rho_\gamma -1) \rho_\nu  \rho_\mu  \delta (\omega_\gamma -\omega_\nu -\omega_\mu -\epsilon )\right.
\nonumber\\
&-\left.[\rho_\alpha  \rho_\gamma  (\rho_\nu +\rho_\mu -1)-\rho_\alpha  \rho_\nu  \rho_\mu 
+(\rho_\gamma -1) \rho_\nu  \rho_\mu ]
e^{-\mathrm{i} t (\omega_\alpha -\omega_\gamma +\omega_\nu +\omega_\mu )} 
\delta \left(\epsilon -\frac{\omega_\alpha +\omega_\gamma -\omega_\nu -\omega_\mu}{2}\right) \right\}.
 \label{eq:broadening_2}
\end{align}
The superscript ``$(2')$'' in  the l.h.s. of Eq.~\eqref{eq:broadening_2} serves to distinguish it from the TDHF contribution \eqref{eq:broadening_hf}.

The contributions \eqref{eq:broadening_2} and \eqref{eq:broadening_hf} have rather similar general structure. Moreover,  
separating in Eq. \eqref{eq:broadening_2} the contribution of matrix elements with only three distinct labels, like those in 
Eq. \eqref{eq:broadening_hf}, one gets an expression whose form is almost identical to  Eq.~\eqref{eq:broadening_hf}, see \ref{sec:qp_broadening}.
This helps to understand how a cancelation required to restore the MBL may work.  Only when taken together, the TDHF and non-TDHF terms describe fully the quantum-coherent dynamics, and thus localization at strong disorder. Discarding some of the terms leads to decoherence and, as a result, to delocalization. Of course, such compensation should be operative to all orders in $V$ to ensure the MBL. We thus expect that if the self-energy in the equation of motion is calculated not to the order $V$ (as in the TDHF) but to the order $V^2$ (or any finite order $V^k$), the system will still experience delocalization for arbitrarily strong disorder at sufficiently long times.

Finally, we note that, since the TDHF approximation destabilizes localization by inducing a finite exponent $\beta$ of anomalous diffusion in the MBL phase, $W > W_c$, it leads, by continuity, to an enhancement of $\beta$ also in the ergodic phase, $W < W_c$---at least, in some vicinity of $W_c$. This is indeed what is observed in the lower panel of Fig.~\ref{fig:comparison_tdvp} (where $W_c \approx 5.5$). It is interesting to note that the performance of the TDHF approximation in quantitative determination of $\beta$ appears to be rather good for moderately strong disorder. For example, for $W=2$, the exact (TDVP) result for $\beta(t)$ as determined in the time range $[50,100]$ is $\beta\approx 0.18$, while the TDHF yields in the same time interval $\beta \approx 0.20$.  It remains to be understood whether there is a regime in which the TDHF framework would provide a parametrically controlled approximation for the quantum dynamics of many-body delocalization.

\section{Summary}

In this paper, we have studied the dynamics of many-body delocalization within the TDHF approximation. Our study included 1D random, 1D quasi-periodic, and 2D random systems. We have accessed large system sizes and very long times $t$ (up to 100 sites and $t=10^5$ for 1D systems and up to $20 \times 20$ sites and $t=10^4$ for 2D systems).  
We have analyzed the dynamics using different initial conditions: in addition to the setting with initial charge-density wave that is particularly popular in experiments and numerical studies, we have also investigated the domain wall melting as well as the dynamics starting from an initial state with alternating occupation of non-interacting states in the energy space. 
Our key results are as follows.

\begin{enumerate}

\item
For all the settings, we have characterized the dynamical process by a running exponent $\beta(t)$. We have demonstrated that this is a very useful characterization of the quantum dynamics that allows one to distinguish sensitively between distinct regimes of many-body delocalization dynamics.

\item
While the TDHF method is \emph{a priori} approximate, it does capture various key properties of the dynamics in highly excited states of interacting disordered (or quasi-periodic) systems. A comparison with the (essentially exact) results by TDVP on large systems (and at relatively short times accessible to TDVP) shows that TDHF performs rather well on the ergodic side of the MBL transition, yielding values of $\beta(t)$ reasonably close to exact ones, see Fig.~\ref{fig:comparison_tdvp}. 

\item
For random 1D systems (Sec.~\ref{spatial-imbalance-1D}, \ref{sec:first_moment}, and \ref{energy-imbalance-1D}), we have found slow, subdiffusive dynamics of many-body delocalization, with the exponent $\beta(t)$ saturating at long times at a disorder-dependent value $\beta$ corresponding to power-law relaxation $\sim t^{-\beta}$. In the whole range of considered disorder strength, the resulting values of this exponent satisfy $\beta < 0.3$, i.e., they are well below the diffusive value 0.5.  We have found consistent values of $\beta$ for all three types of initial conditions, for which we investigated the time evolution of the real-space imbalance, the domain-wall width, and the energy-space imbalance, respectively. These results provide clear evidence of slow, subdiffusive transport on the ergodic side of the MBL transition, thus confirming and extending the corresponding findings of Ref.~\cite{paper:hf_original}. The origin of this slow dynamics is attributed to rare-event Griffiths physics, which is further supported by the study of quasi-periodic systems and of 2D systems (see the next two items in this list).

\item
For quasi-periodic (Aubry-Andr\'e) 1D systems (Sec.~\ref{subsec:quasiperiodic_system}), we have found that $\beta(t)$  characterizing the imbalance has a very different behavior: it increases with time up to its saturation at the ballistic value $\beta=1$. This confirms that the subdiffusive saturation value of $\beta$ in random systems is related to rare localized spots (which are absent in a quasi-periodic system). 

Our findings for the quasi-periodic systems are consistent with previous numerical  results in Refs.~\cite{paper:hf_original, Doggen2019a} where signatures of a faster-than-power-law decay at intermediate times were found. In fact, one of the earliest experimental studies of MBL considered precisely the interacting Aubry-Andr\'e model \cite{Schreiber2015a}. Experimentally, a slow decay of the imbalance was observed. The time range available to the experiment was, however, relatively short; at such times the difference between truly random and quasi-periodic systems is not so pronounced yet. 

\item
For 2D random systems (Sec.~\ref{subsec:two_d}), the imbalance decay is found to follow, at intermediate times, the  $\exp(- \gamma \ln^2 t)$ law (i.e., $\beta (t)$ increases as $\ln t$), as expected within the Griffiths-type picture. At longest times, our results indicate a saturation at $ \beta =1$, which is the value expected from the memory-effect coupling between the large-wave-vector imbalance mode and the diffusive 2D mode. 

Also in the 2D case, experiments find slow transport \cite{Choi2016a} (rather than diffusive behavior). Again, this is not surprising, taking into account not so long times that have been probed experimentally. In such a limited time window (say, $t \sim 10^2$), it is difficult to distinguish the $\exp(- \gamma \ln^2 t)$ decay from a simple power law. This is also the case for TDVP study \cite{paper:elmer_2d} where the characteristic time scale is comparable to that in the experiment. It is a key advantage of the TDHF approach that it provides access to much longer times and thus allows one to see the qualitative difference between 1D and 2D models.

\item While the TDHF approximation characterizes successfully the quantum dynamics in the delocalized phase, there is no true MBL transition in this approximation. Our conclusion in this respect is at variance with that of Ref.~\cite{paper:hf_original} and is consistent with Ref.~\cite{nandy2020dephasing}. We find that $\beta(t)$ characterizing the TDHF data slowly increases with time even for very strong disorder, implying no localization. We have provided also analytical arguments (using an expansion of equations of motion in interaction strength) that shed light on the corresponding mechanism (Sec.~\ref{sec:analytical}). In particular, we have demonstrated that terms providing dephasing occur, within the TDHF approximation, in the second-order in interaction. These terms are very similar in structure to terms of the same order that are discarded by the TDHF approximation (those resulting from the second-order self-energy). We argued that only when both TDHF and non-TDHF terms are retained, the evolution equations correspond to a time-independent Hamiltonian and thus should yield MBL at strong disorder. Keeping only TDHF terms generates dephasing also in the regime of strong disorder where the exact solution would give MBL. 

\end{enumerate}

In conclusion, the TDHF approximation is a very valuable computational tool (complementary to other existing methods) for the analysis of quantum dynamics of disordered and quasi-periodic interacting many-body systems. Crucially, it allows one to explore the wealth of regimes of many-body delocalization dynamics  in large systems and, at the same, at very long time scales. 

\section{Acknowledgments}

We acknowledge useful discussions with S.~Bera, F.~Evers, and M.~Knap. We are particularly thankful for fruitful discussions with I.~Gornyi, especially for his advice on the analytical approach. This research was financially supported by the DFG-RFBR Grant [No. MI 658/12-1 (DFG) and No. 20-52-12034 (RFBR)].

\appendix
\section{Integration step size}
\label{app:integration_step_size}
Solving the TDHF equations of motion~\eqref{eq:hf_eom_same_time} numerically, we have to specify the step size \(\Delta t\) for the integration procedure. All of our results were integrated by steps of \(\Delta t=\unit[10^{-2}]{}\). In this Appendix, we analyze the dependence of the results on the step size.

\begin{figure}[H]
	\centering
\includegraphics[width=\linewidth]{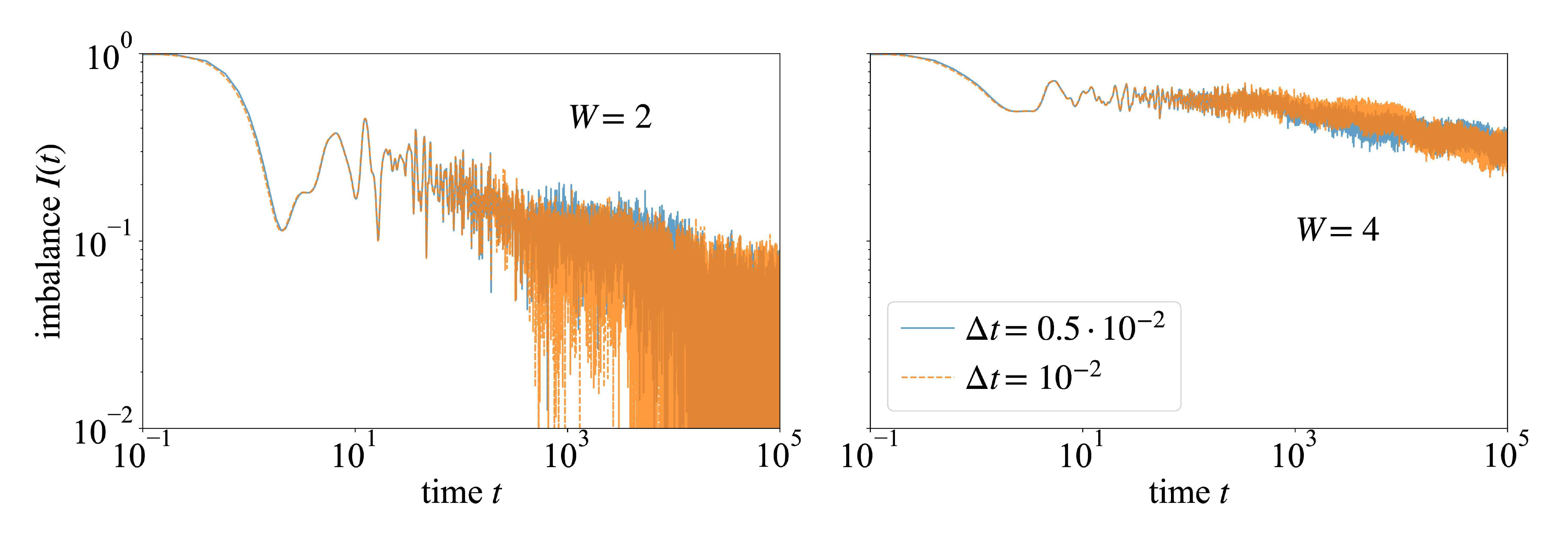}
	\caption{Imbalance in the random 1D model as a function of time for two different step sizes, $\Delta t=\unit[10^{-2}]{}$ and $\Delta t = \unit[0.5 \cdot  10^{-2}]{}$, calculated for the same single disorder realization. The system size is \(L=\unit[50]{}\) sites, with open boundary conditions. The disorder strength is $W=2$ (left panel) and $W=4$  (right panel). Other parameters are \(J=\unit[0.5]{}\) and \(U=\unit[0.5]{}\), as  in Sec.~\ref{subsec:1d_random}. }
	\label{fig:stepsize_check_one_real}
\end{figure}

In Fig.~\ref{fig:stepsize_check_one_real}, we show  a comparison of imbalances in the random 1D model at $W=2$ and $W=4$, as calculated for a single disorder realization with our standard step size \(\Delta t=\unit[10^{-2}]{}\) and smaller steps \(\Delta t = \unit[0.5 \cdot  10^{-2}]{}\) for a system of \(L=\unit[50]{}\) sites, open boundary conditions, and our usual choice of hopping and interaction parameters, \(J=U =\unit[0.5]{}\). In both cases of $\Delta t=\unit[10^{-2}]{}$ and $\Delta t = \unit[0.5 \cdot  10^{-2}]{}$, the same realization of disorder is chosen.

We observe that the imbalance values obtained for simulations with different time steps begin to deviate significantly at \(t \approx \unit[8 \cdot 10^2]{}\) hopping times. The reason for this are fast fluctuations of $I(t)$ for an individual disorder realization superimposed on a smooth decay of $I(t)$. Even though errors originating from a finite time steps are small, they accumulate by the time $t \sim 10^3$ and strongly disturb the specific pattern of these fast fluctuations. 

\begin{figure}[H]
	\centering
	\includegraphics[width=\linewidth]{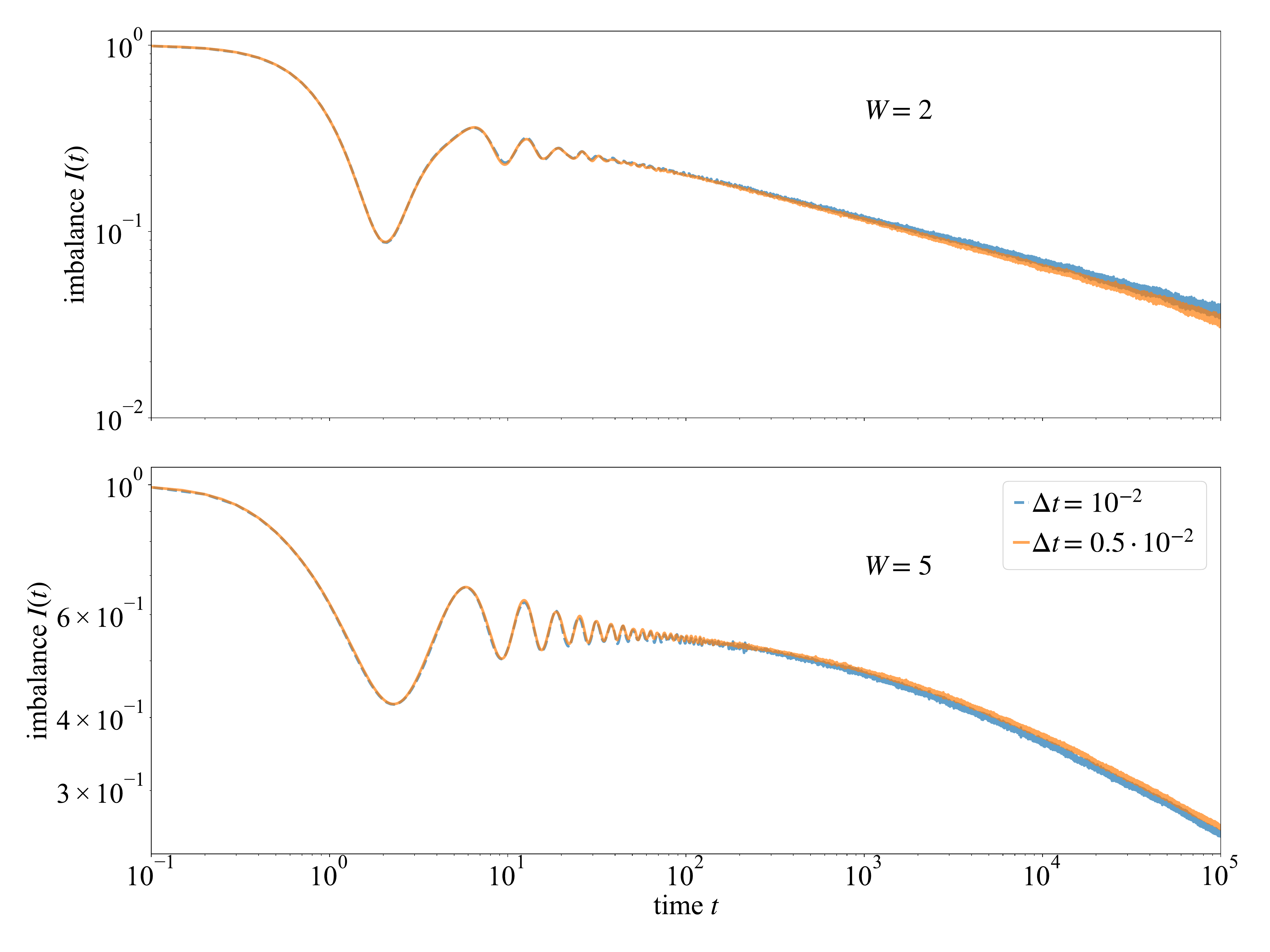}
	\caption{Imbalance in the random 1D model as a function of time for two different step sizes, $\Delta t=\unit[10^{-2}]{}$ and $\Delta t = \unit[0.5 \cdot  10^{-2}]{}$,  averaged over $\approx 500$ disorder realizations. All other parameters are the same as in Fig.~\ref{fig:stepsize_check_one_real}. The system size is \(L=\unit[50]{}\) sites, with open boundary conditions. The disorder strength is $W=2$ (upper panel) and $W=5$  (lower panel). The hopping and interaction parameters are \(J=\unit[0.5]{}\) and \(U=\unit[0.5]{}\), as  in Sec.~\ref{subsec:1d_random}.
	}
	\label{fig:stepsize_check_average}
\end{figure}

We stress, however, that our analysis is based upon averaged quantities: we are interested in the smooth behavior of the imbalance (and other observables), with exact details of fast fluctuations in a given realization of disorder being of no importance. In Fig.~\ref{fig:stepsize_check_average} we present the imbalance averaged over $\approx 500$ realizations of disorder; all other parameters are the same as in 
Fig.~\ref{fig:stepsize_check_one_real}. We see that, upon averaging, the results are nearly identical for two different time steps up to $t = 10^5$, which justifies the analysis of the long-time regime with the chosen step size. Deviations between the traces may serve as a measure of an expected error resulting from a finite step size.

\section{Delocalization by TDHF at strong disorder }
\label{appendix:strong-disorder}

\begin{figure}[H]
	\centering
	\includegraphics[width=\linewidth]{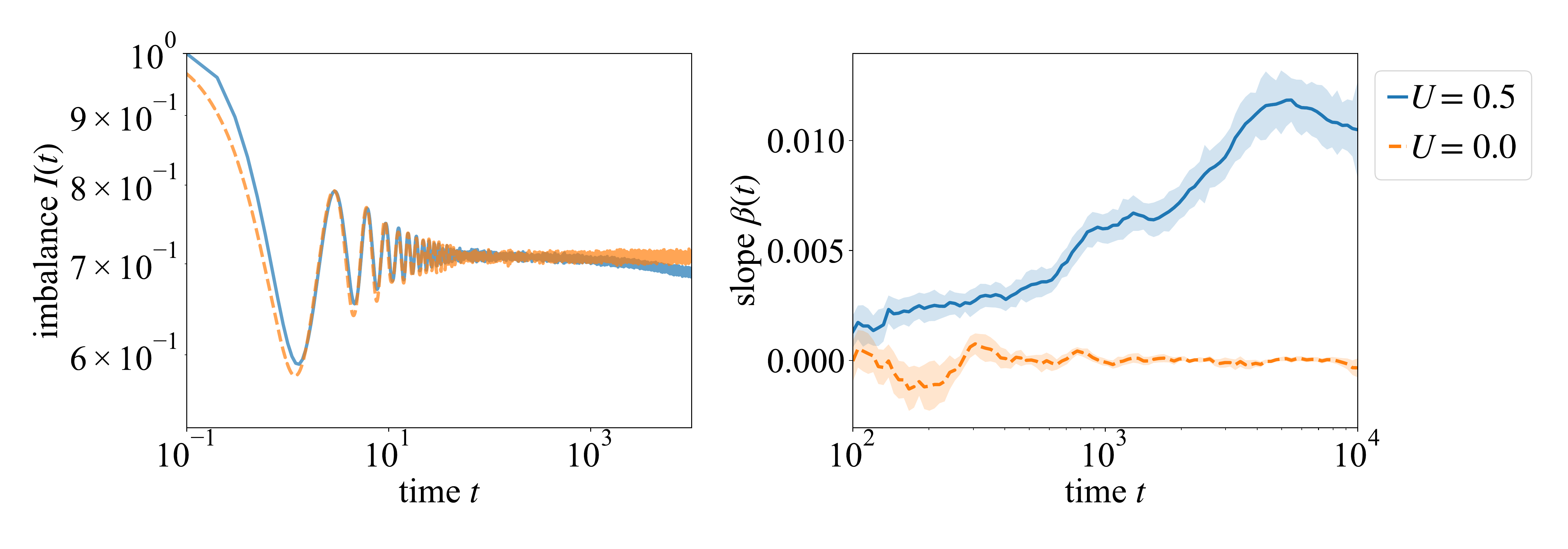}
	\caption{Imbalance dynamics in the random 1D model for \(J=1\), \(W=17\), system size \(L=192\), and periodic boundary conditions. The \(U=0.5\) curve (solid, blue) corresponds to the \(W=17\) TDHF curve in Fig.~1 from Ref.~\cite{paper:hf_original} (averaged over \(\sim 300\) disorder configurations). The \(U=0\) curve (dashed, orange) is the exact noninteracting result (averaged over \(\sim 100\) disorder configurations). The left panel is the imbalance $I(t)$, and the right panel is the corresponding exponent $\beta(t)$.
	}
	\label{fig:comparison_wgk}
\end{figure}

In this Appendix, we provide further evidence that TDHF leads to delocalization even at strong disorder $W$ (in the limit of large time $t$). Specifically, 
we consider the parameters  $J=1$,  $U=0.5$, and $W=17$ as in  Ref.~\cite{paper:hf_original}. The value $W=17$  was the strongest disorder considered in Ref.~\cite{paper:hf_original}; it was argued there that the system is in the localized phase at this disorder within the TDHF.  Indeed, Fig.~1 in Ref.~\cite{paper:hf_original} shows that the dynamics is very slow (which is not surprising at such strong disorder) and might suggest that it is totally frozen. To inspect the imbalance dynamics for these parameters in the TDHF approximation, we plot the corresponding  curve in the left panel of Fig.~\ref{fig:comparison_wgk} in comparison with the exact data for a non-interacting system (for which the imbalance is constant at long $t$ due to localization). We see that the imbalance of the interacting system does show a visible decay. 
This decay is seen in a particularly clear way in the  right panel of Fig.~\ref{fig:comparison_wgk} where the running exponent $\beta(t)$ is shown. The observed relaxation for the interacting model is significantly different from the $\beta=0$ behaviour (corresponding to $I = \text{const}$) of the noninteracting imbalance.

\section{Quasiparticle Broadening}
\label{sec:qp_broadening}

This Appendix presents technical details behind Sec.~\ref{sec:analytical} on the mechanism of dephasing (quasiparticle broadening) that governs many-body delocalization within the TDHF approximation. 
In \ref{sec:qp_hf}, we calculate the TDHF Green's functions and self-energies to the second order in the interaction strength $V$, thus providing the derivation of Eq.~\eqref{eq:broadening_hf}.  
At strong disorder, one expects MBL to result from the exact time evolution of the interacting problem. Thus, the terms in the total Green's function that result from TDHF approximation should combine with those discarded by the TDHF in such a way that the MBL is restored (i.e., no level broadening is generated). To understand how this ``cancellation'' can happen, in \ref{sec:qp_second} we derive the second-order terms in the Green's function that are beyond the TDHF. The result is given by Eq.~\eqref{eq:broadening_2} and shows that TDHF and non-TDHF contributions have indeed a similar structure. A truncation (like that performed within the TDHF approximation), when only a part of the terms is kept, apparently leads to decoherence and thus delocalization at arbitrarily strong disorder, since the quantum dynamics then does not correspond to a time-independent many-body Hamiltonian. 

\paragraph{Keldysh equations of motion: Notations}
In the numerics described in the main text, we have considered the TDHF equations of motion for the Green's function \eqref{eq:hf_eom}. As these equations are nonlinear in $G$, an exact analytical solution within the TDHF scheme turns out to be impossible; however, the locality of the TDHF approximation in time allows for efficient numerical solutions.  
We start with the exact equations of motion, including memory integrals that are non-local in time:
\begin{align}
\mathrm{i}\partial_t G^<(t,t') &= \left[H_0 + \Sigma^{\rm HF}(t)\right]  G^<(t,t') + \int_0^t\dd t^{\prime\prime}\Sigma^R(t,t^{\prime\prime})G^<(t^{\prime\prime},t')- \int_0^{t'}\dd t^{\prime\prime}\Sigma^<(t,t^{\prime\prime})G^A(t^{\prime\prime},t') \,, \label{eq:eom} \\
\mathrm{i}\partial_t G^A(t, t') &= \left[H_0 + \Sigma^{\rm HF}(t)\right]G^A(t, t') - \int_0^{t'}\dd t^{\prime\prime}\Sigma^A(t,t^{\prime\prime})G^A(t^{\prime\prime},t') \,. \label{eq:B2}
\end{align}
Here, we have singled out the Hartree-Fock part of the self-energy,
\begin{align}
\Sigma_{i,j}^{\rm HF}(t)&=  - \mathrm{i} \delta_{i, j} \sum_k V_{i, k} G_{k, k}^< (t, t) 
+ \mathrm{i} V_{i, j} G_{i, j}^<(t, t).
\label{eq:hf_self}
\end{align}
We use Latin indices for position space and Greek indices for the eigenbasis of the non-interacting single-particle Hamiltonian $H_0$.
The remaining part of the self-energy $\Sigma$ is a functional of greater and lesser Green's function $G^>, G^<$ 
of a generic form. This means that, in general, skeleton diagrams of any order (with respect to the interaction $V$) contribute to 
$\Sigma$, starting with the second-order diagrams: 
 $\Sigma^{<}=  \Sigma^{(D_2),<} + \Sigma^{(D_3),<} + \ldots$.  
In particular, the second-order non-HF skeleton diagram yields:
\begin{align}
\Sigma_{i,j}^{(D_2),<}(t,t')&=\sum_{k,l} V_{i,l} V_{j,k} G_{k,l}^>(t',t)\left[G_{l,k}^<(t,t')G_{i,j}^<(t,t')-G_{l,j}^<(t,t')G_{i,k}^<(t,t') \right].
\label{eq:self22}
\end{align}

The retarded, advanced, and Keldysh Green's functions are defined as usual:
\begin{align}
G^R(t,t') &= \Theta(t-t')\left[G^>(t,t')-G^<(t,t')\right],\nonumber\\
G^A(t,t') &= -\Theta(t'-t)\left[G^>(t,t')-G^<(t,t')\right],\nonumber\\
G^K(t,t') &= G^>(t,t')+G^<(t,t').  \label{eq:rak}
\end{align}
Retarded, advanced, and Keldysh self-energies are defined analogously.
The Wigner transform $\tilde{G}(t,\epsilon)$ is defined as
\begin{align}
\tilde{G}(t,\epsilon) &= \int_{-\infty}^{\infty} d\tau e^{\mathrm{i}\epsilon\tau} G(t+\tau/2,t-\tau/2). 
\label{eq:wt}
\end{align}
It is used to determine properties of quasi-particles at time $t$.
The time dependence of $G(t,t)$ can be obtained by integrating $\tilde{G}(t,\epsilon)$ over all energies $\epsilon$.

A convenient for the perturbative expansion notation for the interaction matrix elements is obtained by splitting the interaction strength 
$V$ out of the potential $V_{i,k} = V(\delta_{i,k+1}+\delta_{i,k-1})$ that mediates the density-density interaction.
Specifically, we introduce matrix elements $\mathcal{M}$ written in terms of the exact wavefunctions $\phi_\alpha(i)$ of the noninteracting disordered Hamiltonian $H_0$:
\begin{align}
V\cdot \mathcal{M}^{\alpha \beta}_{\gamma\delta} &\equiv \sum_{i,k} V_{i,k}\mathcal{B}^*_{\alpha\delta}(i,k)\mathcal{B}_{\gamma\beta}(i,k),\\
\mathcal{B}_{\alpha\beta}(i,j) &\equiv  \phi_\alpha(i)\phi_\beta(j)-\phi_\alpha(j)\phi_\beta(i).
\label{def-B}
\end{align}
We choose our wavefunctions (Anderson model) to be real.

\subsection{Hartree-Fock Green's functions}
\label{sec:qp_hf}

In this section, we analyze the Green's functions in the TDHF approximation, discarding the non-Hartree-Fock contributions 
to the self-energy.
The Hartree-Fock time evolution can be mapped to the problem of a non-interacting Hamiltonian $H_0$ supplemented by a time-dependent external field $S(t)$ with an additional constraint that $S(t) = \Sigma^{\mathrm{HF}}[G](t)$.
In the interaction picture, we write:
\begin{align}
\mathrm{i}\partial_t G_I(t,t') &= \Sigma_I(t) G_I(t,t'), \nonumber\\
G_I(t,t') &= \exp(\mathrm{i}H_0t)G(t,t')\exp(-\mathrm{i}H_0t'), \qquad
\Sigma_I(t)= \exp(\mathrm{i}H_0t)\,S(t)\,\exp(-\mathrm{i}H_0t), \nonumber\\
G_I(0,0) &= \mathcal{G}(\rho_0). 
\label{eq:B9}
\end{align}
Here $\mathcal{G}(\rho_0)$ encodes the dependence of the initial condition on the initial density matrix $\rho_0$. 
The solution is given by a formal power series in terms of the time-dependent field $S(t)$:
\begin{align}
G_I(t,t') &= U(t,0)\mathcal{G}(\rho_0)U(0,t'), \qquad
U(t,t') = \mathcal{T} \exp \left(-\mathrm{i}\int_{t'}^{t}dt'' \Sigma_I(t'')\right). 
\label{eq:exp_ansatz}.
\end{align}
If we choose the external field $S(t)$ to be the Hartree-Fock self-energy $\Sigma^{\rm HF}$ defined in \eqref{eq:hf_self}, we get a complicated self-consistent equation. We then perform a perturbative expansion of $G$ and $\Sigma^{\rm HF}$ in powers of $V$:
\begin{align}
G &= G^{(0)} + V G^{(1)} + V^2 G^{(2)} + \mathcal{O}(V^3),
\qquad 
\Sigma^{\rm HF}= V \Sigma^{(1)} + V^2 \Sigma^{(2)} + \mathcal{O}(V^3).
\end{align}
We know $G^{(0)}$ from the free problem (time evolution with $H_0$). With this knowledge, we can compute $\Sigma^{(1)}$ using Eq.~\eqref{eq:hf_self}, which in turn enables us to find $G^{(1)}$ via  Eq.\eqref{eq:exp_ansatz}. Repeating these steps we can go, in principle, to any order in $V$. We carry out now this procedure explicitly up to the second order.

At zeroth order in $V$, we have the Green's functions of a non-interacting Hamiltonian $H_0$ (with eigenenergies $\omega_\alpha$), which we write in the basis of exact single-particle eigenstates:
\begin{align}
\tilde{G}_{\alpha\beta}(t,\epsilon) &
= \int_{-\infty}^{\infty} d\tau \: e^{\mathrm{i}(\epsilon-(\omega_\alpha+\omega_\beta)/2)\tau} 
G_{\alpha\beta}(t+0\mathrm{sign}(\tau),t),
\nonumber\\
\tilde{G}^{(0),R}_{\alpha\beta}(t,\epsilon)&
= \dfrac{\delta_{\alpha\beta}}{\epsilon-(\omega_\alpha+\omega_\beta)/2 + \mathrm{i}0},
\\
\tilde{G}^{(0),<}_{\alpha\beta}(t,\epsilon)&=  
2\pi  \mathrm{i} \: \rho_{0,\alpha\beta} \: \delta\left(\epsilon-\frac{\omega_\alpha+\omega_\beta}{2}\right)\:
e^{-\mathrm{i}(\omega_\alpha-\omega_\beta) t}, 
\label{eq:zero}
\\
\tilde{G}^{(0),>}_{\alpha\beta}(t,\epsilon)&=  
-2\pi \mathrm{i} \: (\delta_{\alpha\beta}-\rho_{0,\alpha\beta}) \:
 \delta\left(\epsilon-\frac{\omega_\alpha+\omega_\beta}{2}\right)\:
 e^{-\mathrm{i}(\omega_\alpha-\omega_\beta) t},
\\
\tilde{G}^{(0),K}_{\alpha\beta}(t,\epsilon)&= 
- 2\pi  \mathrm{i} \: (\delta_{\alpha\beta}-2\rho_{0,\alpha\beta})\:  \delta\left(\epsilon-\frac{\omega_\alpha+\omega_\beta}{2}\right)
\:
e^{-\mathrm{i}(\omega_\alpha-\omega_\beta) t}.
\end{align}
The zeroth order Green's is a Hermitean matrix even if $\rho_0$ does not commute with $H_0$. 
When we start from a diagonal $\rho_0$, e.g., a staggered state in energy space, it is obvious that the Green's function does not explicitly depend on $t$ and the spectrum stays the same.

At higher orders in $V$, for simplicity, we will restrict our consideration to the case when $\rho_0$ is diagonal, 
$\rho_{0,\alpha\beta} = \delta_{\alpha\beta}\rho_\alpha$, and hence the self-energy $\Sigma^{(1)}_{\alpha\beta}$ 
involves the time-independent free Green's function $G^{(0)}(t,t)$ and 
is thus itself time-independent: 
\begin{equation}
\Sigma^{(1)}_{\alpha\beta}=\sum_\gamma\mathcal{M}^{\alpha\beta}_{\gamma\gamma} \rho_{\gamma}.
\label{sigma1ab}
\end{equation}
Below we analyze $G^{<}$ with the initial condition
$$
\mathcal{G}_{\alpha\beta}(\rho_0)=G^{<}_{\alpha\beta}(0,0)=\mathrm{i}\:\rho_\alpha \: \delta_{\alpha\beta}
$$
in equations of motion (\ref{eq:B9})
 (for the other Green's functions one has to change the initial condition 
of the equations of motion). Expanding Eq.~\eqref{eq:exp_ansatz}, we get:
\begin{align}
\tilde{G}^{(1),<}_{\alpha\beta}(t,\epsilon)&
=\int d\tau\,  e^{\mathrm{i}\epsilon \tau}e^{-\mathrm{i}\omega_\alpha(t+\tau/2)}
e^{\mathrm{i}\omega_\beta(t-\tau/2)} \left[\int_0^{t+\tau/2}d t' \Sigma_{I,\alpha\beta }(t') \rho_{\beta}
- \int_0^{t-\tau/2}d t' \rho_{\alpha} \Sigma_{I,\alpha\beta }(t') \right]
\nonumber \\
&
=\int d\tau\,  e^{\mathrm{i}\epsilon \tau}e^{-\mathrm{i}\omega_\alpha(t+\tau/2)} e^{\mathrm{i}\omega_\beta(t-\tau/2)}  \Sigma^{(1)}_{\alpha\beta }
\left[\dfrac{ e^{\mathrm{i}(\omega_\alpha-\omega_\beta )(t+\tau/2)}-1}{\mathrm{i}(\omega_\alpha-\omega_\beta )} 
 \rho_{\beta}
-
\dfrac{e^{\mathrm{i}(\omega_\alpha-\omega_\beta )(t-\tau/2)}-1}{\mathrm{i}(\omega_\alpha-\omega_\beta )} 
\rho_{\alpha} \right]
\nonumber \\
&\stackrel{\alpha\neq \beta}{=}\dfrac{2\pi\mathrm{i}\: \Sigma^{(1)}_{\alpha\beta }}{\omega_\alpha-\omega_\beta}
\left[\rho_{\beta}\delta(\epsilon-\omega_\beta)-\rho_{\alpha}\delta(\epsilon-\omega_\alpha) 
+e^{-\mathrm{i}(\omega_\alpha-\omega_\beta)t}
	\delta\left(\epsilon-\frac{\omega_\alpha+\omega_\beta}{2}\right)\left(\rho_{\alpha}-\rho_{\beta}\right)\right]. 
\label{eq:1o}
\end{align}
To read off the quasiparticle energies and the lifetimes, the full matrix 
$\tilde{G}^<(t,\epsilon) = \sum_n V^n \tilde{G}^{(n),<}(t,\epsilon)$ 
needs to be diagonalized (at fixed $t,\epsilon$). 
This causes additional mixing of the individual orders in the expansion in $V$: in particular, the second-order correction to the diagonalized Green's function will contain a contribution from the first-order off-diagonal terms from Eq.~\eqref{eq:1o}.
Note that new levels $(\omega_\alpha+\omega_\beta)/2$ are introduced by the mixing, with the quasiparticle weight 
$\propto V$.

To the first order in $V$, the diagonal component of $G^<$ can be directly found from the first line of Eq.~\eqref{eq:1o}:
\begin{align}
\tilde{G}^{(1),<}_{\alpha\alpha}(t,\epsilon)&=\int d\tau \cdot e^{\mathrm{i}\epsilon \tau}
e^{-\mathrm{i}\omega_\alpha\tau}\: \tau\: \Sigma^{(1)}_{\alpha\alpha }\rho_{\alpha}
=2\pi \mathrm{i}\delta'(\epsilon-\omega_\alpha)\Sigma^{(1)}_{\alpha\alpha }\rho_{\alpha}.
\label{G-HF-first-order}
\end{align}
This is just the first term in the formal expansion of the zero-order Green's function \eqref{eq:zero} with a shifted dispersion:
\begin{align}
-2\pi\mathrm{i}\delta\left(\epsilon-\omega_\alpha-V\cdot\Sigma^{(1)}_{\alpha\alpha }\rho_{\alpha}\right) &= 
\tilde{G}^{(0),<}_{\alpha\alpha}(t,\epsilon)+V \cdot \tilde{G}^{(1),<}_{\alpha\alpha}(t,\epsilon)+\ldots \label{eq:shift}
\end{align}
This means that the levels $\omega_\alpha$ are shifted by the $t$-independent term $V\Sigma^{(1)}_{\alpha\alpha} = V\sum_\gamma\mathcal{M}_{\gamma\gamma}^{\alpha\alpha}\rho_{\gamma}$. 
In higher-order contributions to the self-energy, we will also identify terms with the derivative of the exact delta-function, $\delta'(\epsilon-\omega_\alpha)$,  with the level shifts.

At \textit{second order} in $V$ there are several contributions: 
\begin{align}
G^{(2),<}_I(t,t') &=G^{(2,1),<}_I(t,t') +G^{(2,2),<}_I(t,t'),
\\
G^{(2,1),<}_I(t,t') &=  \int_0^t \dd t^{\prime\prime}\Sigma_I[G^{(1),<}(t^{\prime\prime},t^{\prime\prime})]\: \rho_0
-\rho_0\int_0^{t'} \dd t^{\prime\prime}\Sigma_I[G^{(1),<}(t^{\prime\prime},t^{\prime\prime})],
\nonumber\\
G^{(2,2),<}_I(t,t')&=-\mathrm{i}\int_0^t \dd t^{\prime\prime}\: \Sigma_I[G^{(0),<}(t^{\prime\prime},t^{\prime\prime})]\int_0^{t^{\prime\prime}} \dd t^{\prime\prime\prime}\Sigma_I[G^{(0),<}(t^{\prime\prime\prime},t^{\prime\prime\prime})]\: \rho_0
\nonumber\\
&+\mathrm{i}\int_0^t \dd t^{\prime\prime}
\Sigma_I[G^{(0),<}(t^{\prime\prime},t^{\prime\prime})]\: \rho_0
\int_0^{t^{\prime}} \dd t^{\prime\prime\prime} \Sigma_I[G^{(0),<}(t^{\prime\prime\prime},t^{\prime\prime\prime})]
\nonumber\\
&-\mathrm{i}\:\rho_0\int_0^t \dd t^{\prime\prime}\: \Sigma_I[G^{(0),<}(t^{\prime\prime},t^{\prime\prime})]\int_0^{t^{\prime\prime}} \dd t^{\prime\prime\prime}\Sigma_I[G^{(0),<}(t^{\prime\prime\prime},t^{\prime\prime\prime})].
\end{align}
The contribution $G^{(2,1),<}$ to $G^{(2),<}$ is obtained by plugging the first-order correction to the Green's function 
\eqref{G-HF-first-order} into the self energy $\Sigma_I$ in the first-order expansion of the evolution operator. 
The other type of contributions is $G^{(2,2),>}$, where $\Sigma_I[G^{(0),>}]$ enters the second-order term in 
expansion of the evolution operator.

Following this procedure, we first obtain $\Sigma_{I,\alpha\beta}^{(2)}$ from Eqs. (\ref{sigma1ab}) and (\ref{eq:1o}):
\begin{align}
\Sigma_{I,\alpha\beta}^{(2)}(t) &\equiv \Sigma_{I,\alpha\beta}[G^{(1),<}(t,t)](t) 
= \sum_{\mu\nu\gamma}\mathcal{M}^{\alpha\beta}_{\mu\nu} 
\mathcal{M}^{\mu\nu}_{\gamma\gamma}\: \rho_{\gamma}\: \left( \rho_{\nu} - \rho_{\mu}\right) \:
e^{\mathrm{i}(\omega_{\alpha}-\omega_{\beta})t}\:
\dfrac{1-e^{\mathrm{i}(\omega_\mu-\omega_\nu)t}}{\omega_\mu-\omega_\nu}.
\label{sigma-HF-second-order}
\end{align}
For the diagonal element of  $G_{\alpha\alpha}^{<,(2,1)}$ we get:
\begin{align}
G_{\alpha\alpha}^{<,(2,1)}(t+\tau/2,t-\tau/2)&=e^{-\mathrm{i}\omega_\alpha\tau}\int_{t-\tau/2}^{t+\tau/2} \dd t^{\prime\prime} \: 
\Sigma_{I,\alpha\alpha}^{(2)}( t^{\prime\prime}) \: \rho_\alpha
\label{G211}\\  
&=\mathrm{i} e^{-\mathrm{i}\omega_\alpha\tau}\sum_{\mu\nu\gamma}
\mathcal{M}^{\alpha\alpha}_{\mu\nu} \mathcal{M}^{\mu\nu}_{\gamma\gamma}\: \rho_{\alpha}\rho_{\gamma} \:
\dfrac{ \rho_{\mu} - \rho_{\nu}}{\omega_\mu-\omega_\nu}
 \left(\tau- \dfrac{2 e^{\mathrm{i}(\omega_\mu-\omega_\nu)t}\sin\frac{(\omega_\mu-\omega_\nu)\tau}{2}}{\omega_\mu-\omega_\nu}\right).
\nonumber
\end{align}
We see that this diagonal component of the TDHF Green's function is fully determined by the matrix 
elements with only three distinct indices.

Note that at $\tau=0$, the contribution to $G_{\alpha\alpha}^{(2),<}$ from Eq.~(\ref{G211}) vanishes,
which implies that the corresponding contribution to the time dependence of the occupation of state $\alpha$ is absent,
thus indicating that Eq.~(\ref{G211}) does not describe decay processes. In order to elucidate the physical meaning 
of this term, we consider its Wigner transform:
\begin{align}
\tilde{G}_{\alpha\alpha}^{(2,1),<}(t,\epsilon)&
=2\pi \mathrm{i} \sum_{\mu\nu\gamma}\mathcal{M}^{\alpha\alpha}_{\mu\nu} \mathcal{M}^{\mu\nu}_{\gamma\gamma}\:
\rho_{\alpha} \rho_{\gamma} \dfrac{ \rho_{\mu} - \rho_{\nu}}{\omega_\mu-\omega_\nu}
\label{G21}
\\
&\times \left\{\vphantom{\frac12}\delta'(\epsilon-\omega_{\alpha}) + \dfrac{e^{\mathrm{i}(\omega_\mu-\omega_\nu)t}}{\omega_\mu-\omega_\nu}
\left[\delta\left(\epsilon-\omega_{\alpha}+\frac{\omega_\mu-\omega_\nu}{2}\right)
-\delta\left(\epsilon-\omega_{\alpha}-\frac{\omega_\mu-\omega_\nu}{2}\right)\right]\right\}.
\nonumber
\end{align}
The time-independent term with the derivative of the delta-function $\delta'(\epsilon-\omega_{\alpha})$ shifts the quasiparticle energy and will be omitted in what follows. 

The other term $G^{(2,2),<}$ is written in terms of the static self-energy parts $\Sigma^{(1)}$:
\begin{align}
&G^{(2,2),<}_{I,\alpha\beta}(t+\tau/2,t-\tau/2)
=-\mathrm{i}\int_0^{t+\tau/2}dt'\int_0^{t'}dt'' 
\Sigma^{(1)}_{I,\alpha \gamma}(t')\Sigma^{(1)}_{I,\gamma \beta}(t'')\rho_{\beta} 
\nonumber \\ 
&+ \mathrm{i} \int_0^{t+\tau/2}dt'\int_0^{t-\tau/2}dt'' 
\Sigma^{(1)}_{I,\alpha\gamma}(t')\rho_{\gamma}\Sigma^{(1)}_{I,\gamma\beta}(t'')
-\mathrm{i}\int_0^{t-\tau/2}dt'\int_0^{t'}dt'' \rho_{\alpha} 
\Sigma^{(1)}_{I,\alpha\gamma}(t')\Sigma^{(1)}_{I,\gamma\beta}(t''),
\end{align}
where a summation over the index $\gamma$ is assumed.
For the diagonal elements ($\alpha=\beta$) we get the Wigner transform in the form:
\begin{align}
\tilde{G}^{(2,2),>}_{\alpha\alpha}(t,\epsilon)
&=
\dfrac{2\pi \mathrm{i}\Sigma^{(1)}_{\alpha\gamma}\Sigma^{(1)}_{\gamma\alpha}}{\omega_{\alpha\gamma}^2}
\left\{
\left[
2\delta(\epsilon-\omega_\alpha)-2\cos(\omega_{\alpha\gamma}t)
\delta\left(\epsilon-\frac{\omega_\alpha+\omega_\gamma}{2}\right)
-\omega_{\alpha\gamma}\delta^\prime(\epsilon-\omega_\alpha)
\right]
\rho_{\alpha}  
\right.
\nonumber
\\
&
\left.
-\left[
\delta(\epsilon-\omega_\alpha)
-2\cos(\omega_{\alpha\gamma}t)\delta\left(\epsilon-\frac{\omega_\alpha+\omega_\gamma}{2}\right)
+\delta(\epsilon-\omega_\gamma)
\right]
\rho_{\gamma}  
\right\},
\label{eq:hf_broad}
\end{align}
where we have introduced a short-hand notation $\omega_{\alpha\gamma}=\omega_\alpha-\omega_\gamma$.
The terms with $\delta(\epsilon-\omega_\alpha)$ and $\delta^\prime(\epsilon-\omega_\alpha)$
contribute to the quasiparticle weight and the shift of energy, respectively. The quasiparticle broadening can be expected to arise
from the terms that contain other energy levels over which the summation is performed, like the term with $\delta(\epsilon-\omega_\gamma)$. We also note that, in contrast to the contribution $G_{\alpha}^{(2,1),<}$, the integral over $\epsilon$
of Eq.~(\ref{eq:hf_broad}), which yields the Green's function at coinciding time arguments and hence the contribution to the occupation of state $\alpha$, does not vanish and hence the contribution of  Eq.~(\ref{eq:hf_broad}) does lead to the TDHF decay. Keeping only the terms responsible for broadening and with 
$\Sigma^{(1)}_{\alpha\gamma}$ expressed through the matrix elements according to Eq. (\ref{sigma1ab}), we get
Eq.~\eqref{eq:broadening_hf} of the main text.

It is convenient to define auxiliary functions combining matrix elements with the combinations of the density matrices entering 
the correction to the Green's function (\ref{G21}) and (\ref{eq:hf_broad}):
\begin{align}
f_\alpha(z)&\equiv\sum_{\mu\nu\gamma}\mathcal{M}^{\alpha\alpha}_{\mu\nu} \mathcal{M}^{\mu\nu}_{\gamma\gamma}\:
\rho_{\alpha} \rho_{\gamma}\:  \dfrac{ \rho_{\nu} - \rho_{\mu}}{(\omega_\mu-\omega_\nu )^2} \delta(\omega_\mu-\omega_\nu-2z),\nonumber\\
p_\alpha(z)&\equiv\sum_{\mu\nu\gamma}\mathcal{M}^{\alpha\gamma}_{\mu\mu} \mathcal{M}^{\gamma\alpha}_{\nu\nu}\:
\rho_{\alpha} \rho_{\nu} \rho_{\mu}\:  \dfrac{ 1}{(\omega_\alpha-\omega_\gamma )^2} \delta(\omega_\alpha-\omega_\gamma-2z),\nonumber\\
q_\alpha(z)&\equiv\sum_{\mu\nu\gamma}\mathcal{M}^{\alpha\gamma}_{\mu\mu} \mathcal{M}^{\gamma\alpha}_{\nu\nu}\:
\rho_{\gamma} \rho_{\nu} \rho_{\mu}\:  \dfrac{ 1}{(\omega_\alpha-\omega_\gamma )^2} \delta(\omega_\alpha-\omega_\gamma-2z).
\label{falphaz}
\end{align}
The second-order TDHF correction $G_{\alpha\alpha}^{(2,1),<}$ is expressed as an integral
in terms of the function $f_\alpha$:
\begin{align}
\tilde{G}_{\alpha}^{(2,1),<}(t,\epsilon) &= 2\pi\mathrm{i}\int \dd z\: f_\alpha(z) \left\{z\delta'(\epsilon-\omega_{\alpha})-e^{2\mathrm{i} z t} \:[\delta(\epsilon-\omega_{\alpha}+z)-\delta(\epsilon-\omega_{\alpha}-z)]\right\}.
\end{align}
By exchanging $\mu,\nu$ in the sum in Eq.~(\ref{falphaz}), one finds that $f_\alpha$ is odd $f_\alpha(z)=-f_\alpha(-z)$. 
The time-dependent contribution to $\rho_\alpha$ is determined by
\begin{align}
\int \dd z f_\alpha(z)\: \cos(2z t) \: \delta(\epsilon-\omega_{\alpha}+z)= -f_\alpha(\epsilon-\omega_{\alpha}) \cos[2(\epsilon-\omega_{\alpha}) t].
\label{B36}
\end{align}
The integral over $\epsilon$, which gives the time decay of the state $\alpha$ is also zero, as was discussed above. 
The symmetry properties of the term (\ref{B36}) are similar to those of the level-shift contribution with $\delta'(\epsilon-\omega_{\alpha})$. 
We conclude that $G_{\alpha}^{(2,1),<}$ is not responsible to TDHF dephasing at order $V^2$. 

Let us now turn to $G^{(2,2),<}$:
\begin{align}
&\tilde{G}^{(2,2),<}_{\alpha\alpha}(t,\epsilon)=2\pi\mathrm{i}\int \dd z\left\{2\cos(2z t)
\left[p_\alpha(z)-q_\alpha(z)\right]\delta(\epsilon-\omega_\alpha+z) 
+q_\alpha(z)\delta(\epsilon-\omega_\alpha+2z)\right\}+\ldots,
\label{G22}
\end{align}
where ``$\ldots$'' denote the terms that are not related to the broadening.
By definition, functions $p_\alpha(z)$ and $q_\alpha(z)$ have both even and odd components as functions of $z$,
unlike the odd function $f_\alpha(z)$. The even components are responsible for the TDHF quasiparticle broadening.
Equation (\ref{G22}) represents a compact form of Eq.~\eqref{eq:broadening_hf} of the main text.

\subsection{Second-order self-energy}
\label{sec:qp_second}

In this section, we compare the broadening due to TDHF self-energy with the broadening due to the second order self-energy 
that is not included in the TDHF approximation. When putting $G^{(0)}$ into the general formula for 
$\Sigma^{(D_2),<}$, Eq.~\eqref{eq:self2}, one obtains the lowest (second-order) contribution in $V$:
\be
\Sigma^{(D_2),<}_{\alpha\beta}(t,t') = -\sum_{ijkl}\sum_{\mu\nu\gamma} V_{il}V_{jk} \mathcal{B}_{\mu\nu}(l,i)\mathcal{B}^*_{\gamma\alpha}(i,l) \mathcal{B}_{\beta\gamma}(j,k)\mathcal{B}^*_{\mu\nu}(k,j) (1-\rho_{\gamma})\rho_{\mu}\rho_{\nu}e^{\mathrm{i}(\omega_\gamma -\omega_\mu-\omega_\nu)(t-t')} \,,
\ee
which yields, upon Wigner transformation,
\be
\tilde{\Sigma}^{(D_2),<}_{\alpha\beta}(\epsilon,t) = -2\pi\mathrm{i}\sum_{\mu\nu\gamma}\mathcal{M}^{\mu\alpha}_{\nu\gamma} \mathcal{M}^{\beta\nu}_{\gamma\mu} (1-\rho_{\gamma})\rho_{\mu}\rho_{\nu}\delta(\epsilon+\omega_\gamma -\omega_\mu-\omega_\nu) \,.
\label{eq:2os}
\ee

The exponential ansatz \eqref{eq:exp_ansatz} is no longer sufficient because of the memory integrals. The implicit equation to solve is
\begin{align}
\mathrm{i}\partial_t G^{<}(t,t') &= \left(H_0+\Sigma^{\rm HF}(t)\right)G^{<}(t,t') + \Omega(t,t'),\nonumber\\
\Omega(t,t') &= \int_0^t\dd t^{\prime\prime}\Sigma^R(t,t^{\prime\prime})G^<(t^{\prime\prime},t')- \int_0^{t'}\dd t^{\prime\prime}\Sigma^<(t,t^{\prime\prime})G^A(t^{\prime\prime},t').
\label{Omega}
\end{align}
We make the ansatz 
\begin{align}
G^{<}(t,t') = \mathrm{i}U^{\rm HF}(t,0)\rho(t,t')U^{\rm HF}(0,t'),
\end{align}
where $U^{\rm HF}(t,0)$ is the evolution operator with the TDHF self-energy.
With $\rho(t=0,t'=0) = \rho_0$, $\rho(t,t')$ has to satisfy:
\begin{align}
\mathrm{i}\partial_t\rho(t,t') &= U^{\rm HF}(0,t)\Omega(t,t')U^{\rm HF}(t',0).
\end{align}
We are only interested in self-energy terms up to second in $V$. We therefore plug the zero-order GF \eqref{eq:zero} into the general expression for the second-order skeleton diagram \eqref{eq:self2}, which yields: 
\begin{align}
\Sigma^{(D_2),<}_{\alpha\beta}(t,t') &=  \sum_{\mu\nu\gamma}\mathcal{M}^{\alpha\gamma}_{\mu\nu} \mathcal{M}^{\gamma\beta}_{\mu\nu}  (1-\rho_{\gamma})\rho_{\mu}\rho_{\nu}e^{i(\omega_\gamma -\omega_\mu-\omega_\nu)(t-t')} \,, \nonumber\\
\Sigma^{(D_2),<}_{\alpha\beta}(t,t') &= \sum_{\gamma\mu\nu}\mathcal{M}^{\alpha\gamma}_{\mu\nu} \mathcal{M}^{\gamma\beta}_{\mu\nu}  \rho_{\gamma}(1-\rho_{\mu})(1-\rho_{\nu})e^{\mathrm{i}(\omega_\gamma -\omega_\mu-\omega_\nu)(t-t')} \,.
\end{align}
For $U^{\rm HF}$ we can then use the zeroth-order term:
\begin{align}
U^{\mathrm{HF},(0)}_{\alpha\beta}(t,t')&= e^{\mathrm{i}\omega_\alpha (t - t')} \delta_{\alpha\beta}.
\end{align}
We thus have to solve the differential equation:
\begin{align}
\mathrm{i}\partial_t\rho^{(2)}(t,t') &= U^{\mathrm{HF},(0)}(0,t)\Omega^{(2)}(t,t')U^{\mathrm{HF},(0)}(t',0),\nonumber\\
\mathrm{i}\partial_{t'}\rho^{(2)}(t,t') &= U^{\mathrm{HF},(0)}(0,t)\Omega^{\prime,(2)}(t,t')U^{\mathrm{HF},(0)}(t',0).
\end{align}
Here, $\Omega^{(2)}$ is the term of the second order in $V$ in the expansion of $\Omega$ from Eq.~\eqref{Omega}. 
The $t'$ evolution is governed by $\Omega^{\prime,(2)}$:
\begin{align}
\Omega^{(2)}(t,t') &=  \int_0^t\dd t^{\prime\prime}\Sigma^{R,(2)}(t,t^{\prime\prime})G^{<,(0)}(t^{\prime\prime},t')- \int_0^{t'}\dd t^{\prime\prime}\Sigma^{<,(2)}(t,t^{\prime\prime})G^{A,(0)}(t^{\prime\prime},t')\nonumber\\
\Omega^{\prime,(2)}(t,t')&=-\int_{t'}^{\infty}\dd t^{\prime\prime}G^{<,(0)}(t,t^{\prime\prime})\Sigma^{R,(2)}(t^{\prime\prime},t')+ \int_{t}^{\infty}\dd t^{\prime\prime}G^{A,(0)}(t,t^{\prime\prime})\Sigma^{<,(2)}(t^{\prime\prime},t'),
\end{align}
where $\tilde{\Omega}(t,t')=U^{\rm HF}(0,t)\Omega(t,t')U^{\rm HF}(t',0)$ includes the phase from the Hartree-Fock time evolution. The condition $\partial_{t'} \tilde{\Omega}^{(2)}(t,t') = \partial_{t} \tilde{\Omega}^{\prime,(2)}(t,t')$ is satisfied, which means the differential equation is exact and the following integral expression is a solution for $\rho^{(2)}(t,t')$:
\begin{align}
\rho^{(2)}(t,t') &= \int_{0}^{t}\dd t \tilde{\Omega}^{(2)}(t,t') + \int_{0}^{t'}\dd t' \tilde{\Omega}^{\prime,(2)}(t,t')-\int_{0}^{t}\dd t\int_{0}^{t'}\dd t' \partial_{t'}\tilde{\Omega}^{(2)}(t,t').
\end{align}
This leads to the following expression for the diagonal component of the
Wigner-transformed Green's function $\tilde{G}_{\alpha\alpha}^{<,(2')}(t,\epsilon)$:
\begin{align}
&\tilde{G}_{\alpha\alpha}^{<,(2')}(t,\epsilon)=4\pi \mathrm{i}\sum_{\mu\nu\gamma}
\dfrac{\mathcal{M}^{\nu\alpha}_{\mu\gamma} \mathcal{M}^{\alpha\mu}_{\gamma\nu}}
{(\omega_\alpha -\omega_\gamma +\omega_\nu +\omega_\mu )^2}
\left\{(\rho_\gamma -1) \rho_\nu  \rho_\mu  \delta (\omega_\gamma -\omega_\nu -\omega_\mu -\epsilon )\right.
\nonumber\\
&+\left.[-\rho_\alpha  \rho_\gamma  (\rho_\nu +\rho_\mu -1)+\rho_\alpha  \rho_\nu  \rho_\mu 
-(\rho_\gamma -1) \rho_\nu  \rho_\mu ]
e^{-\mathrm{i} t (\omega_\alpha -\omega_\gamma +\omega_\nu +\omega_\mu )} 
\delta \left(\epsilon -\frac{\omega_\alpha +\omega_\gamma -\omega_\nu -\omega_\mu}{2}\right)\right. 
\nonumber\\
& -\left. [-\rho_\alpha  \rho_\gamma  (\rho_\nu +\rho_\mu -1)+\rho_\alpha  \rho_\nu  \rho_\mu]
\delta(\epsilon -\omega_\alpha)
-2 \rho_\alpha    
[\rho_\gamma  (\rho_\nu +\rho_\mu -1)-\rho_\nu  \rho_\mu ]\delta' (\omega_\alpha -\epsilon ) \right\}.
\label{NHFG2-full}
\end{align}
The first two terms here yield the quasiparticle broadening,  cf. Eqs.~\eqref{eq:broadening_2}, \eqref{eq:hf_broad}.
 Keeping only these terms, we get Eq.~\eqref{eq:broadening_2} of the main text.
The last two terms, which are proportional to   $\delta (\epsilon-\omega_\alpha)$ and   $\delta' (\epsilon- \omega_\alpha)$, only influence the quasiparticle weight and energies, respectively.

To compare this result with the second-order correction to the TDHF Green's function derived in  \ref{sec:qp_hf}, we retain in Eq. (\ref{NHFG2-full}) only the matrix elements with three distinct indices. 
By definition the stucture of the matrix elements $\mathcal{M}^{\nu\alpha}_{\alpha\gamma} = -\mathcal{M}^{\alpha\alpha}_{\nu\gamma}$ is the same as in Hartree-Fock case. 
We then define auxiliary functions similar to Eq. \ref{falphaz}:
\begin{align}
r_\alpha(z)&\equiv \sum_{\mu\nu}\mathcal{M}^{\alpha\alpha}_{\mu\nu} \mathcal{M}^{\mu\nu}_{\alpha\alpha}\:
\rho_{\alpha} \:  \dfrac{ \rho_{\nu} - \rho_{\mu}}{(\omega_\mu-\omega_\nu )^2} \delta(\omega_\mu-\omega_\nu-2z),\nonumber \\
h_\alpha(z)&\equiv \sum_{\mu\nu}\mathcal{M}^{\alpha\alpha}_{\mu\nu} \mathcal{M}^{\mu\nu}_{\alpha\alpha}\:
\rho_{\alpha} \:  \dfrac{ 2\rho_{\nu}\rho_{\mu} - (\rho_{\mu}+\rho_{\nu})}{(\omega_\mu-\omega_\nu )^2} \delta(\omega_\mu-\omega_\nu-2z).
\label{galphaz}
\end{align}
Up to different prefactors $g_\alpha$ behaves like $f_\alpha$ from Eq. \eqref{falphaz} for $z>0$. The function $h_\alpha$ is an even function of $z$ and determines the Wigner transform of the second-order self energy in Eq. \eqref{eq:2os}. 

Apart from the structure of density matrices, the resulting expression for $\tilde{G}^{<,(2)}_{\alpha\alpha}(t,\epsilon)$
is very similar to the TDHF self-energy and can also be rewritten with the level-spacing distributions introduced in Eq. \eqref{galphaz}:
\begin{align}
&\tilde{G}^{<,(2')}_{\alpha\alpha}(t,\epsilon)\simeq 4\pi\mathrm{i}\rho_\alpha\!\!\int \dd z \left\{[h_{\alpha}(z)-r_{\alpha}(z)]  
  \delta (\epsilon-\omega_\alpha+2z)
-[h_{\alpha}(z)-\rho_\alpha r_{\alpha}(z)] e^{i t z} \delta (\epsilon-\omega_\alpha +z) \right\} +\ldots,
\label{GNHF}
\end{align}
where ``$\ldots$'' denote the non-broadening terms, as well as the contributions of matrix elements with four distinct indices.
Equation \eqref{GNHF} is the compact version of Eq.~\eqref{eq:broadening_2} of the main text, where only matrix elements with three distinct indices are retained. The structure of Eq. (\ref{GNHF}) is very similar to that of the TDHF contribution given by Eq. (\ref{G22}). We see that, depending on the initial state, the TDHF contribution to the decay may, in principle, be compensated by the contribution
of the non-Hartree-Fock type. 

\bibliography{references.bib}

\end{document}